\documentclass[aps,pra,reprint,amsmath,amssymb,superscriptaddress]{revtex4-2}
\usepackage{amsthm}
\usepackage{amsmath}
\usepackage{tikzit}
\usepackage{graphicx}
\usepackage{bm}
\usepackage[toc,page,titletoc]{appendix}
\usepackage{hyperref}
\usepackage[capitalize]{cleveref}

\usepackage{balance}

\usepackage{braket}

\tikzstyle{none-small}=[fill=none, draw=none, shape=circle, tikzit category=misc, tikzit shape=circle, tikzit fill=none, font={\footnotesize}]
\tikzstyle{none-small-gray}=[fill=none, draw=none, shape=circle, text=gray, tikzit category=misc, tikzit shape=circle, tikzit fill=none, font={\footnotesize}]
\tikzstyle{gate}=[shape=rectangle, text height=1ex, text depth=0.25ex, yshift=0.5mm, fill=white, draw=black, minimum height=3mm, yshift=-0.5mm, minimum width=3mm, font={\footnotesize}, tikzit category=circuit]
\tikzstyle{meter}=[shape=rectangle, text height=1ex, text depth=0.25ex, yshift=0.5mm, fill=white, draw=black, minimum height=3mm, yshift=-0.5mm, minimum width=3mm, font={\footnotesize}, tikzit category=circuit, text width=4.5mm, label={{[shift={(-0.05,-0.49)}]\metersymb}}]
\tikzstyle{big gate}=[shape=rectangle, text height=1.5ex, text depth=0.25ex, yshift=0.5mm, fill=white, draw=black, minimum height=10mm, yshift=-0.5mm, minimum width=5mm, font={\normalsize}, tikzit category=circuit]
\tikzstyle{long gate}=[shape=rectangle, text height=1ex, text depth=0.25ex, yshift=0.5mm, fill=white, draw=black, minimum height=3mm, yshift=-0.5mm, minimum width=5mm, font={\footnotesize}, tikzit category=circuit]
\tikzstyle{Z dot}=[inner sep=0mm, minimum size=2mm, shape=circle, draw=black, fill={rgb,255: red,221; green,255; blue,221}, tikzit category=zx]
\tikzstyle{Z phase dot}=[minimum size=5mm, font={\footnotesize\boldmath}, shape=rectangle, rounded corners=2mm, inner sep=1mm, outer sep=-2mm, scale=0.8, tikzit shape=circle, draw=black, fill={rgb,255: red,221; green,255; blue,221}, tikzit draw=blue, tikzit category=zx]
\tikzstyle{X dot}=[Z dot, shape=circle, draw=black, fill={rgb,255: red,255; green,136; blue,136}, tikzit category=zx]
\tikzstyle{X phase dot}=[Z phase dot, tikzit shape=circle, tikzit draw=blue, fill={rgb,255: red,255; green,136; blue,136}, font={\footnotesize\boldmath}, tikzit category=zx]
\tikzstyle{hadamard}=[fill=yellow, draw=black, shape=rectangle, inner sep=0.6mm, minimum height=1.5mm, minimum width=1.5mm, tikzit category=zx]
\tikzstyle{paulibox}=[fill={rgb,255: red,221; green,221; blue,255}, draw=black, shape=rectangle, inner sep=0.6mm, minimum height=5mm, minimum width=5mm, font={\footnotesize}, text height=1.5ex, text depth=0.25ex, tikzit category=zx]
\tikzstyle{vertex}=[inner sep=0mm, minimum size=1mm, shape=circle, draw=black, fill=black, tikzit category=misc]
\tikzstyle{vertex set}=[inner sep=0mm, minimum size=1mm, shape=circle, draw=black, fill=white, font={\footnotesize\boldmath}, tikzit category=misc]
\tikzstyle{small black dot}=[fill=black, draw=black, shape=circle, inner sep=0pt, minimum width=1.2mm, tikzit category=circuit]
\tikzstyle{cnot ctrl}=[fill=black, draw=black, shape=circle, inner sep=0pt, minimum width=1.2mm, tikzit category=circuit]
\tikzstyle{cnot targ}=[fill=white, draw=white, shape=circle, tikzit category=circuit, label={center:$\oplus$}, inner sep=0pt, minimum width=2.1mm, tikzit fill={rgb,255: red,102; green,204; blue,255}, tikzit draw=black]
\tikzstyle{ket}=[fill=white, draw=black, shape=regular polygon, regular polygon sides=3, regular polygon rotate=-30, scale=0.7, inner sep=1pt, tikzit category=circuit, tikzit shape=rectangle, tikzit fill=green]
\tikzstyle{bra}=[fill=white, draw=black, shape=regular polygon, regular polygon sides=3, regular polygon rotate=30, scale=0.7, inner sep=1pt, tikzit category=circuit, tikzit shape=rectangle, tikzit fill=red]
\tikzstyle{scalar}=[shape=rectangle, text height=1.5ex, text depth=0.25ex, yshift=0.5mm, fill=white, draw=black, minimum height=5mm, yshift=-0.5mm, minimum width=5mm, font={\normalsize}]
\tikzstyle{clabel}=[fill=white, draw=none, shape=rectangle, tikzit fill={rgb,255: red,56; green,255; blue,242}, font={\footnotesize}, inner sep=1pt, tikzit category=labels]
\tikzstyle{empty diagram}=[draw={gray!40!white}, dashed, shape=rectangle, minimum width=1cm, minimum height=1cm, tikzit category=misc]
\tikzstyle{cluster small}=[fill=none, thick, draw={rgb,255: red,0; green,128; blue,128}, shape=circle, tikzit category=misc, tikzit shape=circle, minimum size=1.5mm, inner sep=0.3mm, tikzit fill=white, tikzit draw={rgb,255: red,0; green,128; blue,128}, font={\footnotesize}]
\tikzstyle{cluster}=[fill=none, thick, draw={rgb,255: red,0; green,128; blue,128}, shape=circle, tikzit category=misc, tikzit shape=circle, minimum size=3.5mm, inner sep=0pt, tikzit fill=white, tikzit draw={rgb,255: red,0; green,128; blue,128}, font={\footnotesize}]
\tikzstyle{cluster big}=[fill=none, thick, draw={rgb,255: red,0; green,128; blue,128}, shape=circle, tikzit category=misc, tikzit shape=circle, minimum size=4.5mm, text width=2mm, inner sep=0pt, tikzit fill=white, tikzit draw={rgb,255: red,0; green,128; blue,128}, font={\footnotesize}]
\tikzstyle{Cluster dot}=[Z dot, shape=circle, draw=black, fill={rgb,255: red,228; green,26; blue,28}, tikzit category=zx]
\tikzstyle{Cluster opaque1 dot}=[Z dot, opacity=0.6, shape=circle, draw=black, fill={rgb,255: red,190; green,0; blue,0}, tikzit category=zx]
\tikzstyle{Cluster opaque2 dot}=[Z dot, opacity=0.8, shape=circle, draw=black, fill={rgb,255: red,190; green,0; blue,0}, tikzit category=zx]
\tikzstyle{Blue dot}=[Z dot, shape=circle, draw=black, fill={rgb,255: red,51; green,153; blue,255}, tikzit category=zx]

\tikzstyle{hadamard edge}=[-, dashed, dash pattern=on 0.15pt off 1.5pt, line cap=round, very thick, draw={rgb,255: red,51; green,160; blue,44}]
\tikzstyle{box edge}=[-, dashed, dash pattern=on 2pt off 1.5pt, thick, draw={rgb,255: red,55; green,126; blue,184}]
\tikzstyle{brace edge}=[-, tikzit draw=blue, decorate, decoration={brace,amplitude=1mm,raise=-1mm}]
\tikzstyle{diredge}=[->]
\tikzstyle{double edge}=[-, double, shorten <=-1mm, shorten >=-1mm, double distance=2pt]
\tikzstyle{gray edge}=[-, {gray!100}, thick]
\tikzstyle{pointer edge}=[->, very thick, gray]
\tikzstyle{boldedge}=[-, line width=1.2pt, shorten <=-0.17mm, shorten >=-0.17mm]
\tikzstyle{boldedge red}=[<->, line width=1.4pt, shorten <=-0.17mm, shorten >=-0.17mm, draw=red, tikzit draw=red]
\tikzstyle{boldedge blue}=[<->, line width=1.4pt, shorten <=-0.17mm, shorten >=-0.17mm, draw=blue, tikzit draw=blue]
\tikzstyle{dashed black}=[-, dashed, dash pattern=on 2pt off 0.5pt, thick]

\newtheorem{definition}{Definition}
\newtheorem{theorem}{Theorem}
\usepackage{float}
\makeatletter
\let\newfloat\newfloat@ltx
\makeatother
\usepackage{algorithm}
\usepackage{algorithmicx}
\usepackage{algpseudocode}
\algtext*{EndWhile}
\algtext*{EndIf}
\algtext*{EndFor}
\algtext*{EndProcedure}

\crefname{section}{Sec.}{Secs.}
\crefname{figure}{Fig.}{Figs.}
\crefname{appendix}{App.}{Apps.}

\begin{document}
\title{Adaptive Framework for Failure-Aware Protocols in Fusion-Based Graph-State Generation}

\author{Korbinian~Staudacher}
\email{staudacher@nm.ifi.lmu.de}
\affiliation{Ludwig-Maximilian-Universität München, MNM-Team, 80539 Munich, Germany}
\author{Bhilahari Jeevanesan}
\email{bhilahari.jeevanesan@dlr.de}
\affiliation{Remote Sensing Technology Institute, German Aerospace Center (DLR), 82234 Wessling, Germany}
\author{Tobias~Guggemos}
\email{guggemos@nm.ifi.lmu.de}
\email{tobias.guggemos@dlr.de}
\affiliation{Remote Sensing Technology Institute, German Aerospace Center (DLR), 82234 Wessling, Germany}
\affiliation{Ludwig-Maximilian-Universität München, MNM-Team, 80539 Munich, Germany}
\affiliation{University of Vienna, Faculty of Physics, Vienna Center for Quantum Science and Technology (VCQ),\\1090 Vienna, Austria}

    \begin{abstract}
    	We consider the generation of photonic graph states in a linear optics setting where sequential non-deterministic fusion measurements are used to build large graph states out of small linear clusters and develop a framework to optimize the building process using graph theoretic characterizations of fusion networks. We present graph state generation protocols for linear cluster resource states and Type-I/Type-II fusions which are adaptive to fusion failure, that is, they reuse leftover graph states in the remaining building process. To estimate hardware costs, we interpret our protocols as finite Markov processes. This viewpoint allows to cast the expected number of fusion measurements until success as a {\it first passage} problem. We then deploy a pipeline of polynomial algorithms to optimize arbitrary graph states, extract fusion networks and find beneficial orderings of fusions. We check the effectiveness of these algorithms by verifying that the corresponding {\it mean first passage times} are decreased. 
    	We evaluate our pipeline for different initial resource states and fusion mechanisms with varying success probabilities. 
    	Results show reductions in the fusion overhead by several orders of magnitude when compared to simple repeat until success protocols and by up to $40\%$ when compared to state-of-the-art graph state generation protocols, especially for realistic fusion success probabilities between 50-75\%.
    \end{abstract}
    \maketitle
    \section{Introduction}
    Graph states play an important role as a fundamental resource in several quantum computing schemes~\cite{raussendorf2001one,bartolucci_fusion-based_2021}, as scalable entangled units in quantum networks~\cite{azuma2023quantum} and for error correction schemes \cite{yao2012experimental}. 
    They have been realized experimentally on most leading quantum computing implementations such as trapped-ion systems~\cite{lanyon2013measurement,bell2014experimental}, linear optics platforms~\cite{walther2005experimental,kiesel2005experimental}, superconducting systems~\cite{wang201816,mooney2019entanglement} and neutral-atom devices~\cite{bluvstein2022quantum}.
    For linear optics architectures, the major hurdle in building graph states is the non-interacting nature of photons. 
    Thus entangling operations can only be implemented by using non-linear elements like postselecting on measurement outcomes~\cite{knill2001scheme}. 
    While linear clusters of entangled photons can be produced deterministically from quantum emitters~\cite{lindner2009proposal}, creating arbitrary graph states requires either interacting emitters~\cite{li_photonic_2022} or photonic probabilistic entangling measurements, known as fusions~\cite{browne_resource-efficient_2005,hilaire2023near}. 
    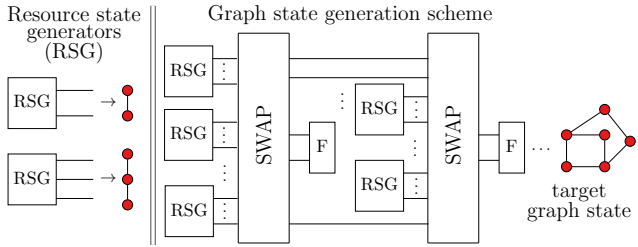
\begin{figure}\label{fig:overview}
    	\resizebox{\linewidth}{!}{
    		\begin{tikzpicture}
	\begin{pgfonlayer}{nodelayer}
		\node [style=big gate] (0) at (-3.85, 3.25) {RSG};
		\node [style=none] (1) at (-2.95, 3.75) {};
		\node [style=none] (2) at (-2.95, 2.75) {};
		\node [style=none] (3) at (-1.95, 3.75) {};
		\node [style=none] (4) at (-1.95, 2.75) {};
		\node [style=none] (5) at (-2.45, 3.45) {$\vdots$};
		\node [style=big gate] (6) at (-3.85, 0.75) {RSG};
		\node [style=none] (7) at (-2.95, 1.25) {};
		\node [style=none] (8) at (-2.95, 0.25) {};
		\node [style=none] (9) at (-1.95, 1.25) {};
		\node [style=none] (10) at (-1.95, 0.25) {};
		\node [style=none] (11) at (-2.45, 0.95) {$\vdots$};
		\node [style=big gate] (18) at (-3.825, -2.25) {RSG};
		\node [style=none] (19) at (-2.925, -1.75) {};
		\node [style=none] (20) at (-2.925, -2.75) {};
		\node [style=none] (21) at (-1.925, -1.75) {};
		\node [style=none] (22) at (-1.925, -2.75) {};
		\node [style=none] (23) at (-2.425, -2.05) {$\vdots$};
		\node [style=none] (24) at (-2.425, -0.55) {$\vdots$};
		\node [style=none] (25) at (-1.9, 4.75) {};
		\node [style=none] (26) at (0.1, 4.75) {};
		\node [style=none] (27) at (-1.9, -3.5) {};
		\node [style=none] (28) at (0.1, -3.5) {};
		\node [style=none] (29) at (-0.9, 0.75) {\large \rotatebox{90}{SWAP}};
		\node [style=none] (30) at (0.1, 0.75) {};
		\node [style=none] (31) at (0.1, -0.25) {};
		\node [style=none] (32) at (0.95, 0.75) {};
		\node [style=none] (33) at (0.95, -0.25) {};
		\node [style=big gate] (34) at (1.4, 0.25) {F};
		\node [style=big gate] (35) at (3.65, 1.75) {RSG};
		\node [style=none] (36) at (4.55, 2.25) {};
		\node [style=none] (37) at (4.55, 1.25) {};
		\node [style=big gate] (38) at (3.65, -1.25) {RSG};
		\node [style=none] (39) at (4.55, -0.75) {};
		\node [style=none] (40) at (4.55, -1.75) {};
		\node [style=none] (41) at (5.55, 2.25) {};
		\node [style=none] (42) at (5.55, 1.25) {};
		\node [style=none] (43) at (5.55, -0.75) {};
		\node [style=none] (44) at (5.55, -1.75) {};
		\node [style=none] (45) at (5.05, 0.45) {$\vdots$};
		\node [style=none] (46) at (5.55, 4.75) {};
		\node [style=none] (47) at (7.55, 4.75) {};
		\node [style=none] (48) at (5.55, -3.5) {};
		\node [style=none] (49) at (7.55, -3.5) {};
		\node [style=none] (50) at (6.55, 0.75) {\large \rotatebox{90}{SWAP}};
		\node [style=none] (51) at (7.55, 0.75) {};
		\node [style=none] (52) at (7.55, -0.25) {};
		\node [style=none] (53) at (8.4, 0.75) {};
		\node [style=none] (54) at (8.4, -0.25) {};
		\node [style=big gate] (55) at (8.85, 0.25) {F};
		\node [style=none] (56) at (0.1, 3.75) {};
		\node [style=none] (57) at (5.575, 3.75) {};
		\node [style=none] (58) at (0.1, 3) {};
		\node [style=none] (59) at (5.575, 3) {};
		\node [style=none] (62) at (0.1, -2.75) {};
		\node [style=none] (63) at (5.575, -2.75) {};
		\node [style=none] (64) at (2.2, 2.3) {$\vdots$};
		\node [style=none] (65) at (10, 0.2) {$\ldots$};
		\node [style=Cluster dot] (66) at (11, -0.5) {};
		\node [style=Cluster dot] (67) at (11, 0.75) {};
		\node [style=Cluster dot] (68) at (12.5, 0.75) {};
		\node [style=Cluster dot] (69) at (12.5, -0.5) {};
		\node [style=Cluster dot] (70) at (12.5, 1.75) {};
		\node [style=Cluster dot] (71) at (13.5, 0.5) {};
		\node [style=big gate] (72) at (-10, 2) {RSG};
		\node [style=none] (73) at (-9.1, 2.5) {};
		\node [style=none] (74) at (-9.1, 1.5) {};
		\node [style=none] (75) at (-7.6, 2.5) {};
		\node [style=none] (76) at (-7.6, 1.5) {};
		\node [style=Cluster dot] (77) at (-6.25, 2.5) {};
		\node [style=Cluster dot] (78) at (-6.25, 1.5) {};
		\node [style=none] (79) at (-7, 2) {$\rightarrow$};
		\node [style=big gate] (80) at (-10, -1) {RSG};
		\node [style=none] (81) at (-9.1, -0.25) {};
		\node [style=none] (82) at (-9.1, -1) {};
		\node [style=none] (83) at (-7.6, -0.25) {};
		\node [style=none] (84) at (-7.6, -1) {};
		\node [style=Cluster dot] (85) at (-6.25, 0) {};
		\node [style=Cluster dot] (86) at (-6.25, -1) {};
		\node [style=none] (87) at (-7, -1) {$\rightarrow$};
		\node [style=none] (89) at (-8.35, 5.5) {\large Resource state};
		\node [style=none] (90) at (-9.1, -1.75) {};
		\node [style=none] (91) at (-7.6, -1.75) {};
		\node [style=Cluster dot] (92) at (-6.25, -2) {};
		\node [style=none] (93) at (-5.25, 5.625) {};
		\node [style=none] (94) at (-5.25, -3.375) {};
		\node [style=none] (95) at (11.5, -1.5) {\large target};
		\node [style=none] (96) at (11.5, -2.25) {\large graph state};
		\node [style=none] (97) at (-8.35, 4.75) {\large generators};
		\node [style=none] (98) at (5.1, 1.95) {$\vdots$};
		\node [style=none] (99) at (5.1, -1.05) {$\vdots$};
		\node [style=none] (100) at (2.5, 5.5) {\large Graph state generation scheme};
		\node [style=none] (101) at (-8.35, 4) {\large (RSG)};
	\end{pgfonlayer}
	\begin{pgfonlayer}{edgelayer}
		\draw (1.center) to (3.center);
		\draw (4.center) to (2.center);
		\draw (7.center) to (9.center);
		\draw (10.center) to (8.center);
		\draw (19.center) to (21.center);
		\draw (22.center) to (20.center);
		\draw (25.center) to (27.center);
		\draw (28.center) to (26.center);
		\draw (26.center) to (25.center);
		\draw (27.center) to (28.center);
		\draw (30.center) to (32.center);
		\draw (33.center) to (31.center);
		\draw (36.center) to (41.center);
		\draw (42.center) to (37.center);
		\draw (39.center) to (43.center);
		\draw (44.center) to (40.center);
		\draw (46.center) to (48.center);
		\draw (49.center) to (47.center);
		\draw (47.center) to (46.center);
		\draw (48.center) to (49.center);
		\draw (51.center) to (53.center);
		\draw (54.center) to (52.center);
		\draw (56.center) to (57.center);
		\draw (58.center) to (59.center);
		\draw (62.center) to (63.center);
		\draw (67) to (68);
		\draw (68) to (69);
		\draw (69) to (66);
		\draw (66) to (67);
		\draw (67) to (70);
		\draw (70) to (71);
		\draw (71) to (69);
		\draw (73.center) to (75.center);
		\draw (76.center) to (74.center);
		\draw (78) to (77);
		\draw (81.center) to (83.center);
		\draw (84.center) to (82.center);
		\draw (86) to (85);
		\draw (91.center) to (90.center);
		\draw (92) to (86);
		\draw [style=double edge] (94.center) to (93.center);
	\end{pgfonlayer}
\end{tikzpicture}
    	}
    	\caption{A schematic overview of the graph state generation process. We consider resource generating modules (RSG), consisting of small entangled units, like 2 or 3-qubit linear cluster states serving as input for a sequential buildup process using fusions (F) on adjacent qubits.}
    \end{figure}

    In this work, we focus on fusion measurements in an all-photonic architecture where we build arbitrary graph states by sequentially fusing small initial resource states (c.f. Fig.~\ref{fig:overview}): At each step we swap qubits and apply a fusion measurement on two adjacent qubits. Based on the fusion outcome, we may then need to add new resource states and again swap qubits and fuse until we have created the desired target graph state. 
    This process of graph state generation can be captured by so-called \textit{fusion networks} which give a configuration of resource states and fusions to reach a target graph state~\cite{bartolucci_fusion-based_2021,lee_graph-theoretical_2023}. 
    In the past years, there has been a surge of effort to optimize graph state generation protocols for photonic architectures including compilation towards fusion networks~\cite{lee_graph-theoretical_2023,Zhang_2023,zilk_compiler_2022} and towards hardware specific emitter schemes~\cite{wein2024minimizing,hilaire2023near,takou2025optimization}. 
    These proposals rely on repeat-until-success schemes with or without redundantly encoded qubits. 
    In other words, if a fusion fails, either the graph states involved in the fusion are built up anew from the beginning or another photon of the redundantly encoded qubit is used to repeat the fusion. 
    
    Here we investigate an {\it adaptive} approach on graph state generation protocols: After a failed fusion we reconfigure the network by incorporating new resource states as well as parts of the already existing graph state.
    We characterize fusion networks by relating fusions to graph-theoretic minor relations and derive procedures to generate fusion networks for arbitrary graphs using either Type-I or Type-II fusion and linear cluster resource states. We then give adaptive protocols to build graph states from fusion networks with strategies to rebuild networks under fusion failure.
    To evaluate the efficiency of such protocols, we view fusion success and failure probabilities in the language of finite Markov processes. 
    This allows us to capture the expected number of fusion attempts until successful graph state generation in terms of the {\it mean first passage time} (MFPT). 
    We set up a scheme to calculate the MFPT from the Markov process and provide algorithms to reduce this time for arbitrary graph states when using our protocols including the search for less costly locally equivalent graph states and optimizing the order in which fusions are applied.
    
    The paper is structured as follows: We start in \cref{sec:prelims} with a discussion of some graph state preliminaries. In \cref{sec:fusion-networks}, we introduce graph-theoretic fusion networks and show how to obtain fusion networks for arbitrary graphs. We introduce adaptive protocols in \cref{sec:graph-state-generation} and show how to evaluate these protocols using mean first passage times in \cref{sec:mft}. We incorporate algorithms to optimize graph state generation towards low mean first passage times in \cref{sec:optimize-fusion-networks} and, finally, evaluate our strategies in \cref{sec:evaluation}.
    
    \section{Graph state preliminaries}\label{sec:prelims}
    Graph states are a family of quantum states where the entanglement structure between qubits can be described as a simple undirected graph $G=(V,E)$, with a vertex set $V$ and edges $E$ ~\cite{hein_multi-party_2004}. The vertices of $V$ of the graph represent the qubits, and entanglement operations are applied between qubits connected by edges in $E$. Formally, a graph state $\ket{G}$, with $G=(V,E)$ being an undirected graph, is of the following form: 
    \begin{equation}
    	\ket{G}=\prod_{(i,j)\in E}\text{CZ}_{i,j}\ket{+}^{\otimes V}
    \end{equation} 
    The CZ operation is a controlled-Z quantum gate which entangles pairs of qubits. Here, they act on qubits in the state ${\ket{+}=(|0\rangle +|1\rangle)/\sqrt{2}}$. Clearly, such graph states are obtained by acting with Clifford gates only, thus they are stabilizer states. An equivalent formulation of this graph state is given by listing its Clifford group generators
    \begin{equation}
         K_i = X_i \prod_{j\in N(i)}Z_j.
    \end{equation} 
    Here, for each qubit $i$ the product extends over all neighbors of $i$, denoted by $N(i)$. Graph states form only a subclass of stabilizer states, yet any stabilizer state can be transformed into a graph state with just local Clifford unitaries~\cite{hein_multi-party_2004}. We next describe some transformations that can be applied to graph states.
    \subsection{Graph theoretic rewrites}\label{sec:graph-rewrites}
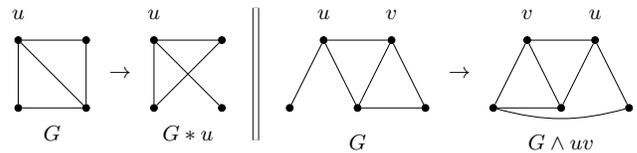
\begin{figure}
    	\centering
        \resizebox{\linewidth}{!}{
    	\begin{tikzpicture}
	\begin{pgfonlayer}{nodelayer}
		\node [style=vertex] (0) at (-9, 2) {};
		\node [style=vertex] (1) at (-7, 2) {};
		\node [style=vertex] (2) at (-9, 0) {};
		\node [style=vertex] (3) at (-7, 0) {};
		\node [style=vertex] (4) at (0, 2) {};
		\node [style=vertex] (5) at (2, 2) {};
		\node [style=vertex] (6) at (-1, 0) {};
		\node [style=vertex] (7) at (1, 0) {};
		\node [style=vertex] (8) at (3, 0) {};
		\node [style=vertex] (9) at (-5, 2) {};
		\node [style=vertex] (10) at (-3, 2) {};
		\node [style=vertex] (11) at (-5, 0) {};
		\node [style=vertex] (12) at (-3, 0) {};
		\node [style=none] (13) at (-6, 1) {$\rightarrow$};
		\node [style=none] (14) at (-2, 2.75) {};
		\node [style=none] (15) at (-2, -0.75) {};
		\node [style=vertex] (16) at (6, 2) {};
		\node [style=vertex] (17) at (8, 2) {};
		\node [style=vertex] (18) at (5, 0) {};
		\node [style=vertex] (19) at (7, 0) {};
		\node [style=vertex] (20) at (9, 0) {};
		\node [style=none] (21) at (4, 1) {$\rightarrow$};
		\node [style=none] (22) at (-9, 2.75) {$u$};
		\node [style=none] (23) at (0, 2.75) {$u$};
		\node [style=none] (24) at (2, 2.75) {$v$};
		\node [style=none] (25) at (6, 2.75) {$v$};
		\node [style=none] (26) at (8, 2.75) {$u$};
		\node [style=none] (27) at (-5, 2.75) {$u$};
		\node [style=none] (28) at (-8, -0.75) {$G$};
		\node [style=none] (29) at (-4, -0.75) {$G\ast u$};
		\node [style=none] (30) at (1, -1) {$G$};
		\node [style=none] (31) at (7, -1) {$G\land uv$};
	\end{pgfonlayer}
	\begin{pgfonlayer}{edgelayer}
		\draw (0) to (1);
		\draw (0) to (3);
		\draw (2) to (3);
		\draw (2) to (0);
		\draw (3) to (1);
		\draw (9) to (10);
		\draw (9) to (12);
		\draw (11) to (9);
		\draw (10) to (11);
		\draw [style=double edge] (14.center) to (15.center);
		\draw (4) to (6);
		\draw (4) to (5);
		\draw (5) to (7);
		\draw (4) to (7);
		\draw (5) to (8);
		\draw (8) to (7);
		\draw (16) to (18);
		\draw (16) to (17);
		\draw (17) to (19);
		\draw (16) to (19);
		\draw (17) to (20);
		\draw [bend left=15] (20) to (18);
		\draw (19) to (18);
	\end{pgfonlayer}
\end{tikzpicture}
        }
    	\caption{Examples of graph-theoretic rewrites on two simple graphs. Left: a local complementation operation is applied at vertex $u$ which toggles the edges between neighbors of $u$. Right: a pivoting operation on edge $\{u,v\}$ toggling edges between exclusive neighbors of $u$, $v$ and the shared neighbors of $u$ and $v$. Further, $u$ and $v$ switch their neighbors.}
    	\label{fig:graph-rewrites}
    \end{figure}

    Given a simple graph $G=(V,E)$ and a vertex $u\in V$, the first graph-rewrite we consider is a so-called \textit{local complementation} on the vertex $u$ that transforms $G$ into a new graph $G\ast u$. This graph is obtained from $G$ by first considering all neighboring vertices $N(u)$ of $u$. When two vertices $v,w\in N(u)$ are connected in $G$, they are disconnected in $G\ast u$ and vice versa. An example is given in \cref{fig:graph-rewrites}. 
    
    An equivalent definition is often given in the literature in terms of the symmetric set difference operator $\vartriangle$. Given two sets $A$ and $B$, this is defined as $A\vartriangle B = {(A\setminus B)\cup (B\setminus A)}$. Thus $A\vartriangle B$ consists of precisely those elements that lie in exactly one of the sets.  
    
    With this, the graph $G\ast u$ has the same vertex set $V$, but the edge set becomes 
    \begin{equation}
        E' = E\vartriangle K_{N(u)},
    \end{equation}
    where $K_{N(u)}$ is the complete graph on vertices $N(u)$ $u$~\cite{kotzig1968eulerian}. Interestingly, another application of $\vartriangle$ will undo the operation, because the set difference operator satisfies
    \begin{equation}
        (A \vartriangle B) \vartriangle B = A.
    \end{equation}

    Another operation that we will apply below is  \textit{pivoting}. This is defined as three alternating local complementations on the endpoints of an edge $\{u,v\}\in E$: 
    \[
    G\land uv=((G\ast u)\ast v)\ast u=((G\ast v)\ast u)\ast v
    \]
    This operation toggles edges between neighboring sets $N(u)\setminus N(v),N(v)\setminus N(u)$ and $N(u)\cap N(v)$ and switches neighbors of $u$ and $v$ (c.f.Fig.~\ref{fig:graph-rewrites}).
    \subsection{Local equivalence}
    A particularly useful property of graph states is that local unitary operations acting on just a single qubit, can be captured via graph theoretic rewrites. Local complementation on a graph state corresponds to the following local Clifford operations (up to a global phase)~\cite{hein_multi-party_2004}: 
    \begin{equation*}
        \ket{G\ast u} = \left[\sqrt{X_u}\prod_{v\in N(u)}\sqrt{Z}_{v}\right]\ket{G}.
    \end{equation*} Graph states are equivalent up to local Clifford operations (LC-equivalent) if and only if their graphs are equivalent up to local complementation~\cite{nest_efficient_2004} and equivalent under local unitary operations (LU-equivalent) if and only if their graphs are equivalent under a generalized $r$-local complementation~\cite{claudet_local_2024}. 
    
    Further, pivoting on a graph state corresponds to the following local operations~\cite{mhalla_graph_2012}: \begin{equation*}
        \ket{G\land uv} = H_uH_vZ_{N(u)\cap N(v)}\ket{G}
    \end{equation*} It has been shown that any real-valued local Clifford equivalence is captured with pivoting~\cite{mhalla_graph_2012}.
    These rewrites are useful in the context of resource minimization in linear optic settings, since local operations are usually easier to realize than entangling operations. To build a graph state on hardware it is often cheaper to build an LU-equivalent graph state with fewer edges first and then transform it to the desired graph state using appropriate local operations~\cite{lee_graph-theoretical_2023}. 

    \section{Fusion networks}\label{sec:fusion-networks}
    In this work, we consider dual-rail encoding, where a qubit is represented as a single photon in two linear optical modes. Fusions are probabilistic measurement operations usually involving two qubits on four modes that serve to glue together different graph states. The pioneering work by Browne and Rudolph~\cite{browne_resource-efficient_2005} introduces two types of fusions based on the detection of one photon for the Type-I fusion gate and two photons for Type-II fusion. These gates are inherently probabilistic, i.e. upon measurement we may also detect a different number of photons. In such cases we say the fusion has failed because the measurement did not create the desired entanglement. The original fusion gates have a success probability of 50\%, yet more recent proposals have shown that both types can be boosted to higher success probabilities using ancilla states~\cite{ewert_34-efficient_2014,bartolucci_creation_2021}. By repeatedly applying fusion measurements we can create large entangled graph states from small initial resource states like linear clusters~\cite{browne_resource-efficient_2005}. 
    Fusion networks~\cite{bartolucci_fusion-based_2021,felice_fusion_2024} define a configuration of fusion operations on graph states which yield a desired target graph state if every fusion succeeds. We first give a general definition in graph-theoretic terms: 
    
    A fusion network is a tuple $(H,F)$ consisting of a simple graph $H=(V,E)$ and a subset $F\subset E$ of edges in $H$, such that $(V,E\setminus F)$ represents a graph state and the endpoints of each edge in $F$ are subject to a two-qubit fusion operation~\footnote{Note, that this definition differs from the definitions given in ~\cite{bartolucci_fusion-based_2021,felice_fusion_2024} where fusions are not represented as separate edges in the graph. We choose this alternate definition since it allows for an easier description of fusion operations as shown in the following sections.}. Given a fusion network $(H,F)$ for a target graph state $\ket{G}$, we say that the fusion process is successful if we reach $G$ from $H$ by applying fusions on all edges in $F$. In all figures of this paper we depict fusion networks as graphs with dashed blue edges for each ``fusion'' edge in $F$ and black edges for each edge $E\setminus F$ belonging to the actual graph state. We give precise definitions of fusion networks for Type-I and Type-II fusion in the next section. 
    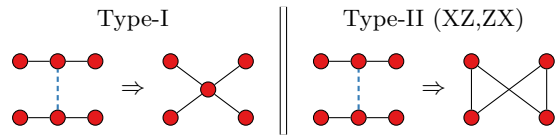
\begin{figure}
    	\begin{center}
    		\begin{tikzpicture}
	\begin{pgfonlayer}{nodelayer}
		\node [style=Cluster dot] (0) at (-7, 0.65) {};
		\node [style=Cluster dot] (1) at (-6, 0.65) {};
		\node [style=Cluster dot] (2) at (-5, 0.65) {};
		\node [style=Cluster dot] (3) at (-7, -0.85) {};
		\node [style=Cluster dot] (4) at (-6, -0.85) {};
		\node [style=Cluster dot] (5) at (-5, -0.85) {};
		\node [style=Cluster dot] (6) at (-3, 0.65) {};
		\node [style=Cluster dot] (8) at (-1, 0.65) {};
		\node [style=Cluster dot] (9) at (-3, -0.85) {};
		\node [style=Cluster dot] (11) at (-1, -0.85) {};
		\node [style=none] (12) at (-4, -0.1) {$\Rightarrow$};
		\node [style=Cluster dot] (13) at (-2, -0.1) {};
		\node [style=Cluster dot] (14) at (1, 0.65) {};
		\node [style=Cluster dot] (15) at (2, 0.65) {};
		\node [style=Cluster dot] (16) at (3, 0.65) {};
		\node [style=Cluster dot] (17) at (1, -0.85) {};
		\node [style=Cluster dot] (18) at (2, -0.85) {};
		\node [style=Cluster dot] (19) at (3, -0.85) {};
		\node [style=Cluster dot] (20) at (5, 0.65) {};
		\node [style=Cluster dot] (21) at (7, 0.65) {};
		\node [style=Cluster dot] (22) at (5, -0.85) {};
		\node [style=Cluster dot] (23) at (7, -0.85) {};
		\node [style=none] (24) at (4, -0.1) {$\Rightarrow$};
		\node [style=none] (26) at (0, 1.9) {};
		\node [style=none] (27) at (0, -1.1) {};
		\node [style=none] (28) at (-4, 1.75) {\small Type-I};
		\node [style=none] (29) at (4, 1.75) {\small Type-II (XZ,ZX)};
	\end{pgfonlayer}
	\begin{pgfonlayer}{edgelayer}
		\draw (0) to (1);
		\draw (1) to (2);
		\draw (3) to (4);
		\draw (4) to (5);
		\draw [style=box edge] (4) to (1);
		\draw (6) to (13);
		\draw (13) to (8);
		\draw (13) to (11);
		\draw (13) to (9);
		\draw (14) to (15);
		\draw (15) to (16);
		\draw (17) to (18);
		\draw (18) to (19);
		\draw [style=box edge] (18) to (15);
		\draw (20) to (23);
		\draw (20) to (22);
		\draw (22) to (21);
		\draw (23) to (21);
		\draw [style=double edge] (26.center) to (27.center);
	\end{pgfonlayer}
\end{tikzpicture}
    	\end{center}
    	\caption{Successful Type-I (left) and Type-II fusion (with observables $\{XZ,ZX\}$) on the middle qubits of two linear clusters with three qubits. The fusion is represented as a dashed blue edge with its endpoints indicating which qubits are inputs of the fusion gate. Upon success, we contract the blue edge in the Type-I case, and apply a Pivot + vertex deletions in the Type-II case.}
    	\label{fig:t1-t2-example}
    \end{figure}
    \subsection{Type-I fusion}
     Upon success, Type-I fusion measures one qubit and leaves the other qubit intact such that it inherits all bonds from the measured qubit (c.f. \cref{fig:t1-t2-example}), whereas upon failure, both qubits are measured in the computational $Z$ basis~\cite{browne_resource-efficient_2005}. We note that in our definition of fusion networks, a successful Type-I fusion $f\in F$ corresponds to contracting the edge $f$, i.e., the two endpoints of the edge are merged into a single vertex~\cite{diestel2025graph}. 
     
     This allows for the following definition of Type-I fusion networks where the notation $H/e$ denotes the graph obtained by contracting the edge $e$ in graph $H$ and $H-S$ the graph obtained by deleting from $H$ all vertices appearing in the vertex set $S$ and all edges incident on $S$:
     \begin{definition} 
     	\label{thm:t-1-fusion-network}
     	Given a fusion network $(H,F)$. A Type-I fusion on two vertices $\{u,v\} \in F$ yields a fusion network $(H',F\setminus \{u,v\})$ with $H'=H/\{u,v\}$ in the success case, that is, we obtain $H'$ from $H$ by contracting the edge $\{u,v\}$. Upon failure we obtain $H'=H - \{u,v\}$.
     \end{definition}
     One can see that the behavior of actual graph state described by $(V,E\setminus F)$ evolves as formalized in~\cite{browne_resource-efficient_2005}.
     We can thus define Type-I fusion networks such that the target graph is obtained from an initial fusion network by repeated edge contractions: 
     \begin{definition}
     	A Type-I fusion network for a graph state $\ket{G}$ is a tuple $(H,F)$ consisting of a simple graph $H$ and a subset $F$ of edges in $H$, such that we obtain $G$ from $H$ by contracting each edge in $F$.
     \end{definition}
     \noindent
     In that sense $G$ can also be seen as graph minor of $H$.
     
     \subsection{Type-II fusion}
     Type-II fusion consumes both qubits and entangles neighboring qubits. There are multiple ways to implement Type-II fusions, yet, most implementations can be described as Bell-state measurements~\cite{bartolucci_creation_2021}. The authors of~\cite{lobl_transforming_2025} distinguish five Bell-state measurements based on different pairs of observables. In this work we focus on Type-II fusion being implemented as measurement in the $\{XZ,ZX\}$ basis, because the effect on the neighborhood of the fused vertices can be described in an accessible graph-theoretic way using the pivot operation. To see this, we look at the stabilizer generators of a graph state after an $\{XZ,ZX\}$ measurement as described in~\cite{lobl_transforming_2025}: 
     \begin{theorem}
     	(Restatement of Equations 8-11 in~\cite{lobl_transforming_2025}) Given a graph state with two non-adjacent qubits $A$ and $B$. Measuring parities $\{X_AZ_B,Z_AX_B\}$ yields the following stabilizer generators:
     	\begin{align}
     		\forall a_i\in N(A)\setminus N(B): X_{a_i}Z_{N(a_i)\vartriangle N(B)} \label{eq:xzzxstablilizers-1}\\
     		\forall b_i\in N(B)\setminus N(A): X_{b_i}Z_{N(b_i)\vartriangle N(A)} \label{eq:xzzxstablilizers-2}\\
     		\forall c_i\in N(A)\cap N(B): X_{c_i}Z_{N(c_i)\vartriangle N(A)\vartriangle N(B)} \label{eq:xzzxstablilizers-3}\\
     		\forall d_i\notin N(A)\cup N(B): X_{d_i}Z_{N(d_i)} 
     	\end{align}
     \end{theorem}
     Recall that in the graph state formalism we get one stabilizer term for each qubit with an $X$ operator acting on the qubit and $Z$ operators acting on all neighboring qubits. 
     From \cref{sec:graph-rewrites}, we know that a pivot toggles edges between three sets $S_1=N(A)\setminus N(B),S_2=N(B)\setminus N(A)$ and $S_3=N(A)\cap N(B)$. Since $S_2\cup S_3=N(B)$, \cref{eq:xzzxstablilizers-1} describes that edges are toggled between $S_1$ and $S_2\cup S_3$. In a similar way \cref{eq:xzzxstablilizers-2,eq:xzzxstablilizers-3} describe that edges are toggled between $S_2$ and $S_1\cup S_3$, resp. $S_3$ and $S_1\cup S_2$.
     This shows that on fusion success the edges between neighbors change exactly as under a pivot operation. With qubits $A$ and $B$ being consumed in the fusion process, we describe the success case as a pivot followed by vertex deletions.
          
     Compared to Type-I fusion, the failure case is more involved since only qubit $u$ gets measured in the $Z$ basis. The other qubit $v$ gets measured in the $X$ basis which has the effect of a pivot on $v$ and a neighbor $w\in N(v) \neq u$ on the graph. This neighbor can be chosen somewhat arbitrarily based on different local transformations~\cite{hein_multi-party_2004}.
     \begin{definition}
     	\label{thm:t-2-fusion-network}
     	Given a fusion network $(H,F)$. A $\{XZ,ZX\}$ fusion on two vertices $\{u,v\}\in F$ yields a fusion network $(H',F\setminus \{u,v\})$ with $H'=(H\land uv) - \{u,v\}$ on success. Upon failure we obtain $H'=((H - \{u\}) \land vw) - \{v\}$ for a neighbor $w\in N(v)$.
     \end{definition}
     We can thus define such a network in terms of pivot minors as introduced in~\cite{mhalla_graph_2012}:
     \begin{definition}
     	An $\{XZ,ZX\}$ fusion network for a graph state $\ket{G}$ is a tuple $(H,F)$ of a simple graph $H$ and a subset $F$ of non-adjacent edges in $H$, such that $G$ is a pivot minor of $H$, i.e., we obtain $G$ from $H$ by applying a pivot on each edge in $F$ and delete its adjacent vertices.
     \end{definition}	
 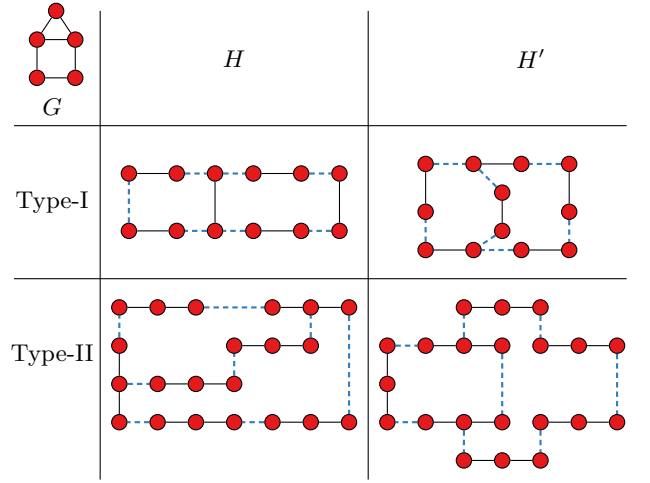
\begin{figure}
     \begin{center}
     \resizebox{\linewidth}{!}{
     	\begin{tikzpicture}
	\begin{pgfonlayer}{nodelayer}
		\node [style=none] (0) at (-4, 7.25) {};
		\node [style=none] (1) at (-4, -5) {};
		\node [style=none] (2) at (-6.25, 4.25) {};
		\node [style=none] (3) at (9.75, 4.25) {};
		\node [style=none] (4) at (-6.25, 0.25) {};
		\node [style=none] (5) at (9.75, 0.25) {};
		\node [style=none] (6) at (3, 7.25) {};
		\node [style=none] (7) at (3, -5) {};
		\node [style=Cluster dot] (50) at (-2, 1.5) {};
		\node [style=Cluster dot] (51) at (-3.25, 1.5) {};
		\node [style=Cluster dot] (52) at (-3.25, 3) {};
		\node [style=Cluster dot] (53) at (-2, 3) {};
		\node [style=Cluster dot] (54) at (-1, 1.5) {};
		\node [style=Cluster dot] (55) at (-1, 3) {};
		\node [style=Cluster dot] (56) at (1.25, 3) {};
		\node [style=Cluster dot] (57) at (0, 3) {};
		\node [style=Cluster dot] (58) at (2.25, 1.5) {};
		\node [style=Cluster dot] (59) at (2.25, 3) {};
		\node [style=Cluster dot] (60) at (0, 1.5) {};
		\node [style=Cluster dot] (61) at (1.25, 1.5) {};
		\node [style=Cluster dot] (62) at (5.75, 1) {};
		\node [style=Cluster dot] (63) at (4.5, 1) {};
		\node [style=Cluster dot] (64) at (4.5, 2) {};
		\node [style=Cluster dot] (65) at (4.5, 3.25) {};
		\node [style=Cluster dot] (66) at (6.5, 1.5) {};
		\node [style=Cluster dot] (67) at (6.5, 2.5) {};
		\node [style=Cluster dot] (68) at (8.25, 2) {};
		\node [style=Cluster dot] (69) at (8.25, 3.25) {};
		\node [style=Cluster dot] (70) at (7, 1) {};
		\node [style=Cluster dot] (71) at (8.25, 1) {};
		\node [style=Cluster dot] (72) at (7, 3.25) {};
		\node [style=Cluster dot] (73) at (5.75, 3.25) {};
		\node [style=none] (74) at (-0.5, 6) {$H$};
		\node [style=none] (75) at (7.25, 6) {$H'$};
		\node [style=Cluster dot] (76) at (-4.65, 6.5) {};
		\node [style=Cluster dot] (77) at (-4.65, 5.5) {};
		\node [style=Cluster dot] (78) at (-5.15, 7.25) {};
		\node [style=Cluster dot] (79) at (-5.65, 6.5) {};
		\node [style=Cluster dot] (80) at (-5.65, 5.5) {};
		\node [style=none] (81) at (-5.25, 4.75) {$G$};
		\node [style=none] (83) at (-5.25, 2.25) {Type-I};
		\node [style=none] (85) at (-5.25, -1.75) {Type-II};
		\node [style=Cluster dot] (86) at (-3.5, -1.5) {};
		\node [style=Cluster dot] (87) at (-3.5, -2.5) {};
		\node [style=Cluster dot] (88) at (-3.5, -3.5) {};
		\node [style=Cluster dot] (89) at (-3.5, -0.5) {};
		\node [style=Cluster dot] (90) at (-2.5, -0.5) {};
		\node [style=Cluster dot] (91) at (-1.5, -0.5) {};
		\node [style=Cluster dot] (92) at (-2.5, -2.5) {};
		\node [style=Cluster dot] (93) at (-1.5, -2.5) {};
		\node [style=Cluster dot] (94) at (-0.5, -2.5) {};
		\node [style=Cluster dot] (95) at (-2.5, -3.5) {};
		\node [style=Cluster dot] (96) at (-1.5, -3.5) {};
		\node [style=Cluster dot] (97) at (-0.5, -3.5) {};
		\node [style=Cluster dot] (98) at (-0.5, -1.5) {};
		\node [style=Cluster dot] (99) at (0.5, -1.5) {};
		\node [style=Cluster dot] (100) at (1.5, -1.5) {};
		\node [style=Cluster dot] (101) at (0.5, -0.5) {};
		\node [style=Cluster dot] (102) at (1.5, -0.5) {};
		\node [style=Cluster dot] (103) at (2.5, -0.5) {};
		\node [style=Cluster dot] (104) at (0.5, -3.5) {};
		\node [style=Cluster dot] (105) at (1.5, -3.5) {};
		\node [style=Cluster dot] (106) at (2.5, -3.5) {};
		\node [style=Cluster dot] (107) at (3.5, -3.5) {};
		\node [style=Cluster dot] (108) at (3.5, -2.5) {};
		\node [style=Cluster dot] (109) at (3.5, -1.5) {};
		\node [style=Cluster dot] (110) at (4.5, -1.5) {};
		\node [style=Cluster dot] (111) at (5.5, -1.5) {};
		\node [style=Cluster dot] (112) at (6.5, -1.5) {};
		\node [style=Cluster dot] (113) at (4.5, -3.5) {};
		\node [style=Cluster dot] (114) at (5.5, -3.5) {};
		\node [style=Cluster dot] (115) at (6.5, -3.5) {};
		\node [style=Cluster dot] (116) at (5.5, -0.5) {};
		\node [style=Cluster dot] (117) at (6.5, -0.5) {};
		\node [style=Cluster dot] (118) at (7.5, -0.5) {};
		\node [style=Cluster dot] (119) at (7.5, -3.5) {};
		\node [style=Cluster dot] (120) at (8.5, -3.5) {};
		\node [style=Cluster dot] (121) at (9.5, -3.5) {};
		\node [style=Cluster dot] (122) at (7.5, -1.5) {};
		\node [style=Cluster dot] (123) at (8.5, -1.5) {};
		\node [style=Cluster dot] (124) at (9.5, -1.5) {};
		\node [style=Cluster dot] (125) at (5.5, -4.5) {};
		\node [style=Cluster dot] (126) at (6.5, -4.5) {};
		\node [style=Cluster dot] (127) at (7.5, -4.5) {};
		\node [style=none] (128) at (6.5, -4.25) {};
	\end{pgfonlayer}
	\begin{pgfonlayer}{edgelayer}
		\draw (0.center) to (1.center);
		\draw (2.center) to (3.center);
		\draw (4.center) to (5.center);
		\draw (6.center) to (7.center);
		\draw (50) to (51);
		\draw (52) to (53);
		\draw (54) to (55);
		\draw (56) to (57);
		\draw (58) to (59);
		\draw (60) to (61);
		\draw [style=box edge] (51) to (52);
		\draw [style=box edge] (53) to (55);
		\draw [style=box edge] (54) to (50);
		\draw [style=box edge] (55) to (57);
		\draw [style=box edge] (56) to (59);
		\draw [style=box edge] (58) to (61);
		\draw [style=box edge] (60) to (54);
		\draw (62) to (63);
		\draw (64) to (65);
		\draw (66) to (67);
		\draw (68) to (69);
		\draw (70) to (71);
		\draw [style=box edge] (63) to (64);
		\draw [style=box edge] (66) to (62);
		\draw [style=box edge] (68) to (71);
		\draw (72) to (73);
		\draw [style=box edge] (65) to (73);
		\draw [style=box edge] (67) to (73);
		\draw [style=box edge] (70) to (62);
		\draw [style=box edge] (69) to (72);
		\draw (76) to (77);
		\draw (78) to (79);
		\draw (79) to (80);
		\draw (76) to (79);
		\draw (80) to (77);
		\draw (76) to (78);
		\draw (86) to (87);
		\draw (87) to (88);
		\draw (89) to (90);
		\draw (91) to (90);
		\draw (92) to (93);
		\draw (94) to (93);
		\draw (95) to (96);
		\draw (97) to (96);
		\draw [style=box edge] (86) to (89);
		\draw [style=box edge] (87) to (92);
		\draw [style=box edge] (95) to (88);
		\draw (98) to (99);
		\draw (100) to (99);
		\draw [style=box edge] (94) to (98);
		\draw (101) to (102);
		\draw (103) to (102);
		\draw [style=box edge] (91) to (101);
		\draw [style=box edge] (100) to (102);
		\draw (104) to (105);
		\draw (106) to (105);
		\draw [style=box edge] (103) to (106);
		\draw [style=box edge] (104) to (97);
		\draw (107) to (108);
		\draw (108) to (109);
		\draw (110) to (111);
		\draw (111) to (112);
		\draw (113) to (114);
		\draw (114) to (115);
		\draw (116) to (117);
		\draw (117) to (118);
		\draw (119) to (120);
		\draw (120) to (121);
		\draw (122) to (123);
		\draw (123) to (124);
		\draw [style=box edge] (109) to (110);
		\draw [style=box edge] (113) to (107);
		\draw (125) to (126);
		\draw (126) to (127);
		\draw [style=box edge] (114) to (125);
		\draw [style=box edge] (111) to (116);
		\draw [style=box edge] (112) to (115);
		\draw [style=box edge] (118) to (122);
		\draw [style=box edge] (124) to (121);
		\draw [style=box edge] (119) to (127);
	\end{pgfonlayer}
\end{tikzpicture}
        }
     \end{center}
     \caption{Different (non-isomorphic) fusion networks $H,H'$ for the same target graph $G$ with either Bell-pairs and Type-I fusion, or 3-qubit linear clusters and Type-II ($\{XZ,ZX\}$) fusion.}
     \label{fig:fusion-networks-combined}
 \end{figure}

    \subsection{Generating fusion networks}\label{sec:generating-fusion-networks}
    
    In this section we consider the problem of finding a suitable fusion network for a given graph state $\ket{G}$ using either two-qubit linear clusters and Type-I fusion or three-qubit linear clusters and Type-II fusion. Note that in general, fusion networks are not unique for a given target graph. As shown in \cref{fig:fusion-networks-combined}, there can be many non-isomorphic networks leading to the same graph.
    
\begin{figure}
    	\begin{center}
        \resizebox{\linewidth}{!}{
    		\begin{tikzpicture}
	\begin{pgfonlayer}{nodelayer}
		\node [style=Cluster dot] (0) at (-4.25, 4.5) {};
		\node [style=Cluster dot] (1) at (-2.75, 4.5) {};
		\node [style=Cluster dot] (2) at (-1.25, 4.5) {};
		\node [style=Cluster dot] (3) at (-10.25, 4.5) {};
		\node [style=Cluster dot] (4) at (-8.75, 4.5) {};
		\node [style=none] (5) at (-4.25, 5.25) {$a$};
		\node [style=none] (6) at (-2.75, 5.3) {$b$};
		\node [style=none] (7) at (-10.25, 5.25) {$a$};
		\node [style=none] (8) at (-8.75, 5.3) {$b$};
		\node [style=Cluster dot] (12) at (-10.5, 1.75) {};
		\node [style=Cluster dot] (13) at (-9, 1.75) {};
		\node [style=none] (16) at (-10.5, 2.5) {$(a)$};
		\node [style=none] (17) at (-9, 2.475) {$a$};
		\node [style=none] (18) at (-11.5, 1.75) {};
		\node [style=Cluster dot] (19) at (-7.5, 1.75) {};
		\node [style=none] (20) at (-7.5, 2.5) {$b$};
		\node [style=Cluster dot] (21) at (-4.75, 2) {};
		\node [style=Cluster dot] (22) at (-3.25, 2) {};
		\node [style=none] (23) at (-4.75, 2.75) {$(a)$};
		\node [style=none] (24) at (-1.75, 2.75) {$a$};
		\node [style=none] (25) at (-5.75, 2) {};
		\node [style=Cluster dot] (26) at (-1.75, 2) {};
		\node [style=none] (27) at (-0.25, 2.85) {$b$};
		\node [style=Cluster dot] (28) at (-0.25, 2) {};
		\node [style=Cluster dot] (29) at (-11.5, -1.25) {};
		\node [style=Cluster dot] (30) at (-10.25, -1.25) {};
		\node [style=none] (31) at (-11.5, -0.45) {$(a)$};
		\node [style=none] (32) at (-10.25, -0.475) {$a$};
		\node [style=none] (33) at (-12.5, -1.25) {};
		\node [style=Cluster dot] (34) at (-8.75, -1.25) {};
		\node [style=none] (35) at (-8.75, -0.4) {$b$};
		\node [style=Cluster dot] (36) at (-4.75, -0.5) {};
		\node [style=Cluster dot] (37) at (-3.25, -0.5) {};
		\node [style=none] (38) at (-4.75, 0.25) {$(a)$};
		\node [style=none] (39) at (-1.75, 0.25) {$a$};
		\node [style=none] (40) at (-5.75, -0.5) {};
		\node [style=Cluster dot] (41) at (-1.75, -0.5) {};
		\node [style=Cluster dot] (43) at (-0.25, -0.5) {};
		\node [style=Cluster dot] (44) at (-7.5, -1.25) {};
		\node [style=none] (45) at (-7.5, -0.45) {$(b)$};
		\node [style=none] (46) at (-6.5, -1.25) {};
		\node [style=Cluster dot] (47) at (-4.75, -2.25) {};
		\node [style=none] (48) at (-4.75, -1.5) {$(b)$};
		\node [style=none] (49) at (-5.75, -2.25) {};
		\node [style=Cluster dot] (50) at (-3.25, -2.25) {};
		\node [style=none] (51) at (-1.75, -1.5) {$b$};
		\node [style=Cluster dot] (52) at (-1.75, -2.25) {};
		\node [style=Cluster dot] (54) at (-0.25, -2.25) {};
		\node [style=none] (57) at (-9.5, 7) {Type-I};
		\node [style=none] (58) at (-2.75, 7) {Type-II};
		\node [style=none] (60) at (-14, 2.5) {$a\in V$};
		\node [style=none] (61) at (-14, -1.75) {$b\in V$};
		\node [style=none] (62) at (-12.75, 8) {};
		\node [style=none] (63) at (-12.75, -3) {};
		\node [style=none] (64) at (-6.25, 8) {};
		\node [style=none] (65) at (-6.25, -3) {};
		\node [style=none] (66) at (-15, 6.25) {};
		\node [style=none] (67) at (0, 6.25) {};
		\node [style=none] (68) at (-14.05, 1.5) {$b\notin V$};
		\node [style=none] (71) at (-10.5, 0.9) {$a$};
		\node [style=none] (75) at (-7.5, 1) {$b$};
		\node [style=none] (76) at (-11.5, 0.9) {or:};
		\node [style=none] (77) at (-13.95, 5.25) {$a\notin V$};
		\node [style=none] (78) at (-14, 4.25) {$b\notin V$};
		\node [style=none] (79) at (-14, -0.75) {$a\in V$};
	\end{pgfonlayer}
	\begin{pgfonlayer}{edgelayer}
		\draw (3) to (4);
		\draw (2) to (1);
		\draw (1) to (0);
		\draw (12) to (18.center);
		\draw [style=box edge] (12) to (13);
		\draw (13) to (19);
		\draw (21) to (25.center);
		\draw [style=box edge] (21) to (22);
		\draw (22) to (26);
		\draw (26) to (28);
		\draw (29) to (33.center);
		\draw [style=box edge] (29) to (30);
		\draw (30) to (34);
		\draw (36) to (40.center);
		\draw [style=box edge] (36) to (37);
		\draw (37) to (41);
		\draw (41) to (43);
		\draw (44) to (46.center);
		\draw [style=box edge] (44) to (34);
		\draw (47) to (49.center);
		\draw (50) to (52);
		\draw (52) to (54);
		\draw [style=box edge] (50) to (47);
		\draw [style=box edge] (43) to (54);
		\draw (62.center) to (63.center);
		\draw (64.center) to (65.center);
		\draw (67.center) to (66.center);
	\end{pgfonlayer}
\end{tikzpicture}
            }
    	\end{center}
    	\caption{Adding an edge $\{a,b\}\in G$ to a fusion network for both Type-I and Type-II networks. Characters in brackets denote the labeling before adding the edge, characters without brackets the updated labeling.}
    	\label{fig:tree-vs-loop-edges}
    \end{figure}
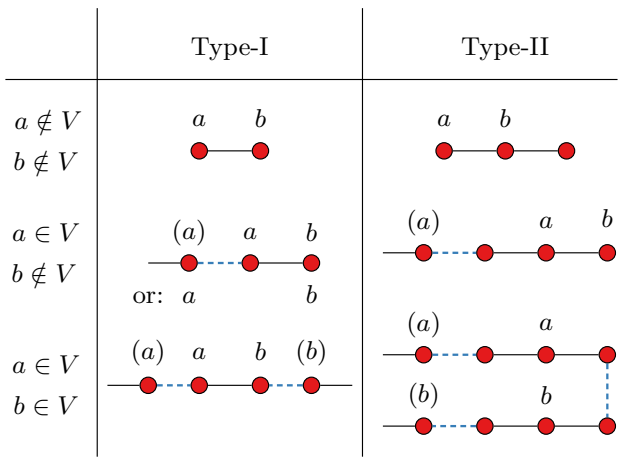

    To construct a fusion network $(H,F)$ for a target graph $G$ we traverse the edges of $G$ and add them sequentially to the fusion network with an internal labeling to keep track of which vertex in $H$ corresponds to which vertex in $G$. Given a simple graph $H=(V,E)$, we distinguish three cases for adding an edge $\{a,b\}$, see \cref{fig:tree-vs-loop-edges}: 
    \begin{description}
    	\item[$\mathbf{a,b\notin V}$] Both endpoints of the edge are not in the fusion network yet. 
        We add a linear cluster to the fusion network and label two adjacent vertices as $a$ and $b$. 
        For Type-II fusion there remains an unlabeled vertex which does not get consumed by a fusion, yet we may use this vertex when appending future edges. 
    	\item[$\mathbf{a\in V;b\notin V}$] Only one of the endpoints is present in the fusion network. 
        We add a linear cluster and fuse one of its endpoints with the existing vertex. 
        For Type-II fusion we label the other two vertices of the linear cluster as our new $a$ and $b$ vertices, for Type-I fusion we can choose which qubit gets measured upon fusion.
    	\item[$\mathbf{a,b\in V}$] Both endpoints are present in the fusion network. 
        Here we need to fuse both vertices of the two-qubit cluster with the endpoints in the Type-I case. 
        For Type-II fusion we add two new clusters, fuse them together to a four-qubit cluster and fuse their endpoints with the existing network.
    \end{description}
    
    We give an illustrative algorithm for constructing Type-I fusion networks in App.~\ref{app:Pseudocode}. The Type-II algorithm is basically the same, however, it needs to manage unused vertices from the 3-qubit linear clusters.
    For Type-I fusion and 2-qubit linear clusters the traversal strategies result in fusion networks with a minimal number of resource states. Since the unmeasured qubit in a successful Type-I fusion just inherits all edges from the measured qubit, it never increases the number of edges in a graph. For generating a graph of $n$ edges with Type-I fusion we therefore need at least $n$ 2-qubit linear clusters and since our algorithm introduces exactly one resource state for each edge it is optimal in that sense. 
    
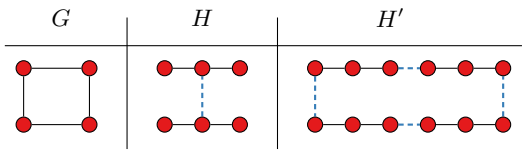
\begin{figure}
    	\begin{center}
    	\begin{tikzpicture}
	\begin{pgfonlayer}{nodelayer}
		\node [style=Cluster dot] (6) at (-0.75, -0.85) {};
		\node [style=Cluster dot] (7) at (1, 0.65) {};
		\node [style=Cluster dot] (8) at (-0.75, 0.65) {};
		\node [style=Cluster dot] (9) at (1, -0.85) {};
		\node [style=none] (27) at (2, 2) {};
		\node [style=none] (28) at (2, -1.5) {};
		\node [style=none] (29) at (-1.25, 1.25) {};
		\node [style=none] (30) at (6, 2) {};
		\node [style=none] (31) at (6, -1.5) {};
		\node [style=none] (32) at (12.5, 1.25) {};
		\node [style=none] (33) at (0.25, 2) {$G$};
		\node [style=none] (34) at (4, 2) {$H$};
		\node [style=none] (35) at (9, 2) {$H'$};
		\node [style=Cluster dot] (36) at (3, 0.65) {};
		\node [style=Cluster dot] (37) at (4, 0.65) {};
		\node [style=Cluster dot] (38) at (5, 0.65) {};
		\node [style=Cluster dot] (39) at (3, -0.85) {};
		\node [style=Cluster dot] (40) at (4, -0.85) {};
		\node [style=Cluster dot] (41) at (5, -0.85) {};
		\node [style=Cluster dot] (42) at (7, 0.65) {};
		\node [style=Cluster dot] (43) at (8, 0.65) {};
		\node [style=Cluster dot] (44) at (9, 0.65) {};
		\node [style=Cluster dot] (45) at (10, 0.65) {};
		\node [style=Cluster dot] (46) at (11, 0.65) {};
		\node [style=Cluster dot] (47) at (12, 0.65) {};
		\node [style=Cluster dot] (48) at (7, -0.85) {};
		\node [style=Cluster dot] (49) at (8, -0.85) {};
		\node [style=Cluster dot] (50) at (9, -0.85) {};
		\node [style=Cluster dot] (51) at (10, -0.85) {};
		\node [style=Cluster dot] (52) at (11, -0.85) {};
		\node [style=Cluster dot] (53) at (12, -0.85) {};
	\end{pgfonlayer}
	\begin{pgfonlayer}{edgelayer}
		\draw (6) to (9);
		\draw (6) to (8);
		\draw (8) to (7);
		\draw (9) to (7);
		\draw (27.center) to (28.center);
		\draw (30.center) to (31.center);
		\draw (32.center) to (29.center);
		\draw (36) to (37);
		\draw (37) to (38);
		\draw (39) to (40);
		\draw (40) to (41);
		\draw [style=box edge] (40) to (37);
		\draw (42) to (43);
		\draw (43) to (44);
		\draw (45) to (46);
		\draw (46) to (47);
		\draw [style=box edge] (44) to (45);
		\draw (48) to (49);
		\draw (49) to (50);
		\draw [style=box edge] (42) to (48);
		\draw (51) to (52);
		\draw (52) to (53);
		\draw [style=box edge] (50) to (51);
		\draw [style=box edge] (53) to (47);
	\end{pgfonlayer}
\end{tikzpicture}

    	\end{center}
    	\caption{Two $\{XZ,ZX\}$ fusion networks $H,H'$ for obtaining the target graph $G$. $H$ needs less resource states and fusions, yet, our fusion network strategies would only find a fusion network similar to $H'$.}
    	\label{fig:inefficient-fusion-network}
    \end{figure}

    Yet, this is not the case for Type-II fusion. Our algorithms always fuse in such way, that at least one endpoint of a resource state is involved, i.e. a node with degree one. Like in the Type-I case this results in fusions never changing the number of edges in the target graph. But a pivot may also increase the number of edges when applied on two vertices with higher degrees (c.f. \cref{fig:inefficient-fusion-network}) which shows that there are smaller Type-II fusion networks with less resource states which are not found by our traversing strategy. Improving these fusion networks for fixed target graphs is possible (c.f.~\cite{wein2024minimizing}), but it is likely that a general algorithmic approach one has to give up small fusion networks and pay the cost of larger computational complexity.

    \section{Adaptive graph state generation}\label{sec:graph-state-generation}
\begin{figure*}
    	\begin{tikzpicture}
	\begin{pgfonlayer}{nodelayer}
		\node [style=Cluster dot] (0) at (-13.75, 3.5) {};
		\node [style=Cluster dot] (1) at (-12.75, 3) {};
		\node [style=Cluster dot] (2) at (-12.75, 4) {};
		\node [style=Cluster dot] (3) at (-11.75, 4) {};
		\node [style=Cluster dot] (4) at (-10.5, 4) {};
		\node [style=Cluster dot] (5) at (-10.5, 3) {};
		\node [style=Cluster dot] (6) at (-11.75, 3) {};
		\node [style=Cluster dot] (7) at (-9.5, 3) {};
		\node [style=Cluster dot] (8) at (-7, 1.5) {};
		\node [style=Cluster dot] (9) at (-7, 0.5) {};
		\node [style=Cluster dot] (10) at (-8, 1) {};
		\node [style=Cluster dot] (11) at (-6, 0.5) {};
		\node [style=none] (14) at (-6, 3) {after failure};
		\node [style=vertex] (15) at (0, 1.5) {};
		\node [style=vertex] (16) at (1, 0.5) {};
		\node [style=vertex] (17) at (1, 1.5) {};
		\node [style=vertex] (18) at (-1, 1.5) {};
		\node [style=vertex] (19) at (0, 0.5) {};
		\node [style=none] (20) at (0, 3) {Subgraph $H'$};
		\node [style=Cluster dot] (22) at (6.5, 1.5) {};
		\node [style=Cluster dot] (23) at (8.75, 0.5) {};
		\node [style=Cluster dot] (24) at (7.75, 0.5) {};
		\node [style=Cluster dot] (25) at (6.5, 0.5) {};
		\node [style=Cluster dot] (26) at (5.5, 1.5) {};
		\node [style=Cluster dot] (27) at (4.5, 1.5) {};
		\node [style=none] (28) at (14.25, 3) {Integration};
		\node [style=Cluster dot] (29) at (10.75, 1) {};
		\node [style=Cluster dot] (30) at (11.75, 0.5) {};
		\node [style=Cluster dot] (31) at (11.75, 1.5) {};
		\node [style=Cluster dot] (32) at (15.75, 1.5) {};
		\node [style=Cluster dot] (33) at (15.75, 0.5) {};
		\node [style=Cluster dot] (34) at (13.75, 1.5) {};
		\node [style=Cluster dot] (35) at (14.75, 0.5) {};
		\node [style=Cluster dot] (36) at (14.75, 1.5) {};
		\node [style=Cluster dot] (37) at (12.75, 1.5) {};
		\node [style=Cluster dot] (38) at (13.75, 0.5) {};
		\node [style=Cluster dot] (39) at (12.75, 0.5) {};
		\node [style=none] (40) at (-3.25, 4.25) {};
		\node [style=none] (41) at (3.25, 4.25) {};
		\node [style=none] (42) at (10, 4.25) {};
		\node [style=none] (43) at (-3.25, -2.75) {};
		\node [style=none] (44) at (3.25, -2.75) {};
		\node [style=none] (45) at (10, -2.75) {};
		\node [style=none] (46) at (-8.75, 4.25) {};
		\node [style=none] (47) at (-8.75, -2.75) {};
		\node [style=none] (48) at (-14.5, 2.25) {};
		\node [style=none] (49) at (18, 2.25) {};
		\node [style=none] (50) at (-14.5, -0.25) {};
		\node [style=none] (51) at (18, -0.25) {};
		\node [style=Cluster dot] (52) at (-7, -1) {};
		\node [style=Cluster dot] (53) at (-7, -2) {};
		\node [style=Cluster dot] (54) at (-8, -1.5) {};
		\node [style=Cluster dot] (55) at (-6, -2) {};
		\node [style=Cluster dot] (56) at (-4, -2) {};
		\node [style=vertex] (57) at (-0.5, -1) {};
		\node [style=vertex] (58) at (1.5, -2) {};
		\node [style=vertex] (59) at (0.5, -2) {};
		\node [style=vertex] (60) at (-0.5, -2) {};
		\node [style=vertex] (61) at (-1.5, -1) {};
		\node [style=vertex] (62) at (-2.5, -1) {};
		\node [style=vertex] (63) at (2.5, -2) {};
		\node [style=Cluster dot] (64) at (6, -1) {};
		\node [style=Cluster dot] (65) at (9.25, -2) {};
		\node [style=Cluster dot] (66) at (8.25, -2) {};
		\node [style=Cluster dot] (67) at (5, -1) {};
		\node [style=Cluster dot] (68) at (6, -2) {};
		\node [style=Cluster dot] (69) at (5, -2) {};
		\node [style=Cluster dot] (70) at (4, -1) {};
		\node [style=Cluster dot] (71) at (7.25, -2) {};
		\node [style=Cluster dot] (72) at (4, -2) {};
		\node [style=Cluster dot] (73) at (10.75, -1.5) {};
		\node [style=Cluster dot] (74) at (11.75, -2) {};
		\node [style=Cluster dot] (75) at (11.75, -1) {};
		\node [style=Cluster dot] (76) at (14.75, -1) {};
		\node [style=Cluster dot] (77) at (14.75, -2) {};
		\node [style=Cluster dot] (78) at (15.75, -2) {};
		\node [style=Cluster dot] (79) at (13.75, -1) {};
		\node [style=Cluster dot] (80) at (15.75, -1) {};
		\node [style=Cluster dot] (81) at (16.75, -1) {};
		\node [style=Cluster dot] (82) at (12.75, -1) {};
		\node [style=Cluster dot] (83) at (16.75, -2) {};
		\node [style=Cluster dot] (84) at (17.75, -1) {};
		\node [style=none] (85) at (-11.75, 1) {Type-I};
		\node [style=none] (86) at (-11.75, -1.5) {Type-II};
		\node [style=Cluster dot] (93) at (-5, 0.5) {};
		\node [style=Cluster dot] (94) at (-4, 0.5) {};
		\node [style=none] (95) at (6.75, 3.75) {Fusion network};
		\node [style=none] (96) at (-11.75, 4.5) {$v_1$};
		\node [style=none] (97) at (-10.5, 4.5) {$v_2$};
		\node [style=none] (98) at (-6, 3.75) {$(H,F)$ };
		\node [style=none] (99) at (6.75, 3) {for $H'$};
	\end{pgfonlayer}
	\begin{pgfonlayer}{edgelayer}
		\draw (0) to (1);
		\draw (3) to (2);
		\draw (2) to (0);
		\draw (1) to (2);
		\draw (5) to (4);
		\draw [style=box edge] (4) to (3);
		\draw (3) to (6);
		\draw (7) to (5);
		\draw (8) to (9);
		\draw (10) to (8);
		\draw (9) to (10);
		\draw (17) to (16);
		\draw (15) to (18);
		\draw (19) to (15);
		\draw (24) to (23);
		\draw (25) to (22);
		\draw (26) to (27);
		\draw [style=box edge] (22) to (26);
		\draw (29) to (30);
		\draw (31) to (29);
		\draw (30) to (31);
		\draw (33) to (32);
		\draw (36) to (35);
		\draw (37) to (34);
		\draw (38) to (39);
		\draw [style=box edge] (34) to (38);
		\draw [style=box edge] (31) to (37);
		\draw [style=box edge] (38) to (35);
		\draw [style=box edge] (36) to (32);
		\draw (42.center) to (45.center);
		\draw (44.center) to (41.center);
		\draw (40.center) to (43.center);
		\draw (46.center) to (47.center);
		\draw (48.center) to (49.center);
		\draw (50.center) to (51.center);
		\draw (52) to (53);
		\draw (54) to (52);
		\draw (53) to (54);
		\draw (59) to (58);
		\draw (57) to (60);
		\draw (61) to (57);
		\draw (62) to (61);
		\draw (58) to (63);
		\draw (66) to (65);
		\draw (67) to (64);
		\draw (68) to (69);
		\draw [style=box edge] (64) to (68);
		\draw (70) to (67);
		\draw (72) to (69);
		\draw (66) to (71);
		\draw (73) to (74);
		\draw (75) to (73);
		\draw (74) to (75);
		\draw (78) to (77);
		\draw (79) to (76);
		\draw (80) to (81);
		\draw [style=box edge] (76) to (80);
		\draw (82) to (79);
		\draw (84) to (81);
		\draw (78) to (83);
		\draw [style=box edge] (75) to (82);
		\draw [style=box edge] (81) to (83);
		\draw (94) to (93);
	\end{pgfonlayer}
\end{tikzpicture}
    	\caption{Example of both Type-I and Type-II rebuild behaviors. The original fusion network is in the upper left corner. After fusion on the dashed blue edge $\{v_1,v_2\}$ failed, we first update the fusion network, then build a subgraph $H'$ for rebuilding destroyed edges and integrate the resulting fusion network for $H'$ into the existing one. For Type-II failure, we consider $v_1$ to be $Z$-measured and $v_2$ to be $X$-measured.}
    	\label{fig:rebuild-t1-t2}
    \end{figure*}
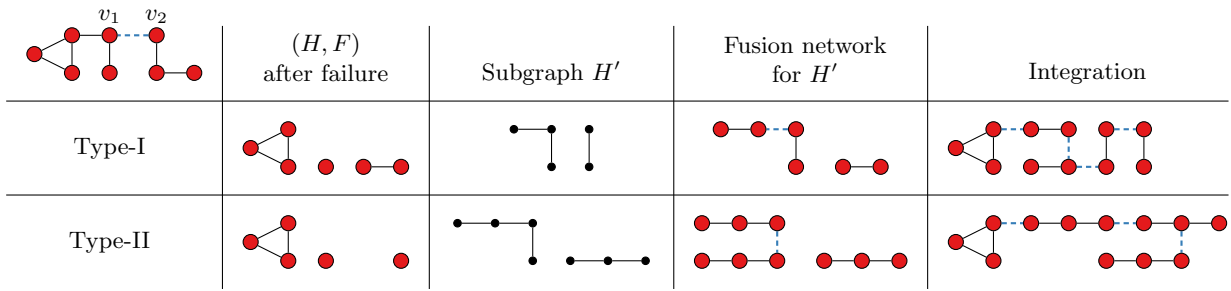

    We now give protocols to build graph states from fusion networks which recycle graph states after a fusion failure. 
    As a general procedure, we start with a fusion network $(H,F)$ describing our initial photonic system, sequentially apply fusions in $F$ and update the fusion network according to the graph-theoretic transformations in \cref{sec:fusion-networks}. In case of fusion failure, we rebuild the network by combining the remaining parts of the graph state with new resource states using the following steps: \begin{enumerate}
    	\item We remove some vertices in $H$ which got modified by fusion failure. As a result,  $H$ may become disconnected, decomposing into multiple connected components.
    	\item We create a graph $H'$ modeling the graph structure of the removed vertices. 
    	\item We add additional edges to $H'$ where its vertices connect to remaining graph components of $H$.
    	\item We build a fusion network for $H'$ and use the additional edges to fuse it with $H$.
    \end{enumerate}
    After the rebuild procedure, we obtain a new fusion network for the target graph state. We illustrate the rebuild protocol for Type-I and Type-II fusions in \cref{fig:rebuild-t1-t2}.

    For the Type-I fusion protocol we again consider 2-qubit linear clusters as resource states. Fusion failure on an edge $\{v_1,v_2\}\in F$ performs a $Z$ measurement on both qubits, thereby removing them from the graph state. The subgraph $H'$ thus consists initially only of vertices $v_1$ and $v_2$. We then iterate over all connected components $C$ in $H$ and add edges to $H'$ for each edge connecting $v_1$ or $v_2$ with $C$ before fusion failure. It can happen that $C$ contains just a single qubit. In this case, we discard the component and just add a new edge to $H'$ since we assume that inserting a new 2-qubit linear cluster resource state is cheaper than rebuilding the network with the leftover component. We then use the algorithm from \cref{sec:generating-fusion-networks} to build a fusion network for $H'$ and integrate it into $H$. A pseudocode algorithm for the entire Type-I protocol can be found in App.~\ref{app:Pseudocode}.
    
    For Type-II fusion, we use 3-qubit linear clusters as resource states. Here, fusion failure measures one vertex in $Z$ and the other in $X$. As an $X$ measurement also affects the entanglement between neighbors of the measured vertex, we cannot just rebuild all edges involving $v_1$ and $v_2$ to obtain a valid fusion network. To circumvent this problem, we discard the neighbors of the $X$ measured vertex and thus create a subgraph $H'$ containing vertices $v_1$, $v_2$ and all neighbors of the $X$ measured vertex. 
    As a convention, we choose the vertex with the least neighbors to receive an $X$ measurement to make rebuilding less costly~\footnote{Note, that convention can also be adapted to hardware by configuring the linear optical elements in a Type-II fusion (c.f.~\cite{lobl_transforming_2025})}. 
    We iterate over connected components just as we did in the Type-I case with the distinction that we add two edges to the subgraph for each original edge we want to rebuild. Further, we now discard a component if it contains less than $3$ qubits since we have three-qubit linear resource states available.

    \section{Markov Processes and First Passage Times}\label{sec:mft}
    In this section, we develop a framework for evaluating the efficiency of fusion networks and fusion orders by calculating the {\it average time} until the attempted fusions lead, after several steps, to the desired target graph state for the first time. The protocols described in the previous section define a {\it stochastic discrete Markov process} on a state space with a finite dimension that we denote by $N$. 
    
    At each step a fusion operation succeeds with probability $p$ and fails with probability $q = 1-p$. The state space of the Markov process consists of all intermediate stages of the graph state following the protocol and is fully captured by a $N\times N$ transition matrix $P$ encoding all possible transitions between states with their respective probabilities. More precisely, $P_{ij}$ is equal to the probability to transition to state $i$ in the next step, if the system is currently in state $j$. We denote this probability also by $p_{i\leftarrow j}$. Note that $P$ is in general not a symmetric matrix since transitions are not necessarily reversible in one step. However, since a transition to \textit{any} state, possibly the same state, must occur with probability $1$, $P$ is a stochastic matrix with columns that all add up to $1$, i.e: we have $\sum_i{P_{ij}} = 1$.
    
    Let $\mathbf{\upsilon}$ now denote a vector containing the occupation probabilities in state space, that is, the system is found in state $i$ with probability $\mathbf{\upsilon}_i$. After one step of the Markov process the new occupation probabilities are given by $P\mathbf{\upsilon}$. The late time occupation probabilities are found by applying $P$ repeatedly to the initial state $\mathbf{\upsilon}_0$. Since all entries of $P$ are non-negative, there exists an eigenvector $\mathbf{\pi}$ with all positive entries and largest eigenvalue equal to $1$ \cite{kemeny1969finite}. Under generic conditions (irreducibility and aperiodicity) only this eigenmode survives in the late-time limit. Thus $\mathbf{\pi}$ is the stationary distribution: \begin{equation}
    	\lim_{n\rightarrow\infty}P^n\mathbf{\upsilon}_0=\mathbf{\pi}
    \end{equation}
    
\begin{figure}
    	\begin{center}
    		\resizebox{.8\linewidth}{!}{
    			\input{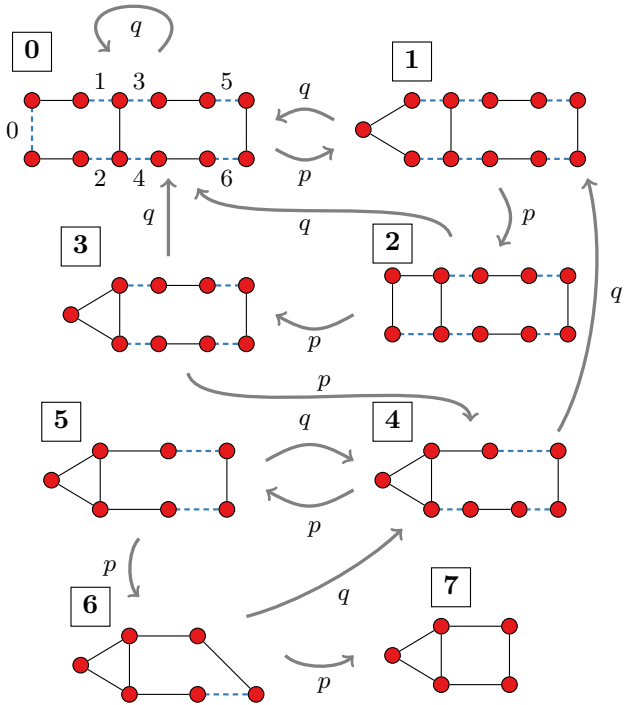}
    		}
    	\end{center}
    	\caption{Exemplary transition graph for creating a 6-qubit graph state \fbox{8} from two-qubit linear clusters. The order in which the Type-I fusions are carried out is indicated next to the edges in state \fbox{0}.}
    	\label{fig:transitions}
    \end{figure}

    To illustrate the construction of a transition matrix, we consider the creation of the 5-qubit graph state shown in \cref{fig:transitions}. 
    We start from the initial resource state labeled configuration \fbox{0}. The Markov process ends when we eventually reach the finished graph state shown as state \fbox{7}. The transition matrix $P$ capturing all possible transitions between the states in \cref{fig:transitions} is given by \[P=
        \begin{pmatrix}
        	q & p & 0 & 0 & 0 & 0 & 0 & 0 & 0 \\
        	q & 0 & p & 0 & 0 & 0 & 0 & 0 & 0 \\
        	q & 0 & 0 & p & 0 & 0 & 0 & 0 & 0 \\
        	0 & 0 & 0 & 0 & p & q & 0 & 0 & 0 \\
        	0 & 0 & 0 & 0 & 0 & q & p & 0 & 0 \\
        	0 & q & 0 & p & 0 & 0 & 0 & 0 & 0 \\
        	0 & 0 & 0 & 0 & 0 & 0 & 0 & q & p \\
        	0 & q & 0 & 0 & p & 0 & 0 & 0 & 0 \\
        	0 & 0 & 0 & 0 & 0 & 0 & 0 & 0 & 1 \\
        \end{pmatrix}
    \]
    
    With the underlying Markov process clearly defined, we can now focus on the main quantity that is of interest to us,  the so-called \textit{Mean First-Passage-Time} (MFPT). This is the average number of steps that it takes to reach the target state for the first time, when starting from an initial state. This problem has been well studied in the literature, see ~\cite{kemeny1969finite,redner2001guide} for pedagogical introductions. Here, we confine ourselves to a self-contained introduction to computing the MFPT.
    
     Let the matrix $M_{ij} = M_{i\leftarrow j}$ denote the MFPT to start in state $j$ and arrive in state $i$ for the first time. We want to set up a recursive relation for $M_{i\leftarrow j}$ that will allow us to compute its value. To this end we argue as follows: Reaching $i$ from $j$ may be possible directly in one step, or alternatively through an intermediate state $k\neq i$. The probability for the former process is $p_{i\leftarrow j}$. The latter process results in a total number of steps $1+M_{i\leftarrow k}$ and occurs with probability $p_{k\leftarrow j}$, thus we have the recursive relation \begin{equation}
    	M_{i\leftarrow j} = p_{i\leftarrow j} + \sum_{k\neq i}p_{k\leftarrow j} (1+M_{i\leftarrow k})
    \end{equation}
    This allows us to derive the following relation between the transition matrix $P$ and the MFPT matrix $M$: \begin{equation}
    	M = E + M\cdot P - D\cdot P
    \end{equation} where $E$ is a matrix with all entries equal to $1$ and $D$ is a diagonal matrix with entries \begin{equation}
    D_{ij} = \frac{\delta_{ij}}{\pi_j}.
\end{equation}
    We provide a pedagogical derivation of this result in \cref{app:FirstPassage}. A straightforward inversion of this relation is not possible, since $\mathbf{1}-P$ is a singular matrix. To see this, note that the columns of $P$ all sum to $1$. Thus, the all-one vector $(1,1,\ldots,1)$ is a left eigenvector of $\mathbf{1}-P$ with eigenvalue zero which means $\mathbf{1}-P$ is not invertible. In \cref{app:FirstPassage}, we show that $M$ is instead given by the following expression
    \begin{equation}
    	M = D(\mathbf{1}-Z+Z_0E),
    \end{equation}
    where $Z$ is called the fundamental matrix and is defined as \begin{equation}\label{eq:Zinversion}
    	Z = (\mathbf{1} - P + \Pi)^{-1}.
    \end{equation}
    The matrix $Z_0$ is the diagonal matrix of $Z$, i.e: $(Z_0)_{ii} = Z_{ii}$ and $(Z_0)_{ij} = 0$ for $i\neq j$ and the matrix $\Pi$ consists of $N$ repeated columns of the $\pi$ vector. The derivation of these results is found in  \cref{app:FirstPassage}. 
    
    To summarize, from the knowledge of the transition matrix $P$, we can calculate the MFPT matrix $M$ explicitly. This will be heavily utilized below to assess the quality of different fusion schemes.

    We emphasize that while our pipeline presented in the following sections consists of algorithms with polynomial runtime, the MFPT calculation itself has an exponential runtime. Thus we do not include the latter in the pipeline and instead only use the MFPT to benchmark small instances of fusion networks. 

    Nevertheless, we point out that the Markov chain is always finite. Its starting state is a fusion network of $N$ qubits. Every state with $n$ qubits in the fusion network results in $n-1$ qubits (Type I) or $n-2$ qubits (Type 2) in case of success. In case of fusion failure, the fusion network has at most $N$ qubits. If a rebuild requires more than $N$ qubits, we start anew from a fusion network with $N$ qubits. Thus the total number of qubits is always limited to at most $N$. This is the worst case node size of our network. 
    Thus our Markov chain is definitely finite and upper bounded, in the worst case, by the total number $T_N$ of unlabeled graphs on $N$ nodes. This number is easily calculated since we have $N(N-1)/2$ edges in the graph and each can be present or not. Thus there are 
    \begin{eqnarray}
    T_N = \frac{2^{N(N-1)/2}}{N!}
    \end{eqnarray}
    different unlabeled graphs. The denominator takes care of overcounting by dividing by the $N!$ permutations of the node labels.  This is the worst case size of the Markov chain. 

    The building of the Markov chain and construction of the sparse transition matrix requires time $\mathcal{O}(T_N)$. The inversion in Eq.~\eqref{eq:Zinversion} by Gaussian Elimination requires $\mathcal{O}(T_N^3)$ operations. The latter dominates the total runtime. 
    
    \section{Optimizing fusion networks}\label{sec:optimize-fusion-networks}
    With mean-first passage times as tool to evaluate fusion network protocols we now describe algorithms to optimize the graph state generation process at different stages.
    \subsection{Finding locally equivalent graphs with less edges}
    A first step towards minimizing mean-first passage times is to find LU-equivalent graph states requiring less entangling operations. Since our network generation algorithms require at least one fusion per edge, reducing the number of edges in the target graph directly reduces the number of fusions. We then build this graph state and transform it to the desired graph state with local unitary operations which are comparably much cheaper than fusions.
    
    Given a graph state, finding an LU-equivalent one with minimum edges seems to be a difficult problem even when restricting the unitaries to Clifford gates. If a graph is LC-equivalent to a tree, we can find a sequence of local complementations to transform the graph into the tree in polynomial time~\cite{bouchetTrees}. Since local complementation preserves connectivity and we assume our initial graph to be connected, there can be no LC-equivalent graph with fewer edges than a tree. Yet, we are usually interested in graphs not equivalent to trees since the latter can be generated deterministically by a single quantum emitter~\cite{lindner2009proposal}. For the general case, there are strong indications that the edge minimization problem is NP-complete~\cite{sharma_minimizing_2025}. 
    
    Therefore, given an arbitrary graph we first check whether the graph is equivalent to a tree using the algorithm in~\cite{bouchetTrees} and if not, we use a greedy approach to minimize edges with local complementation as introduced in~\cite{staudacher_reducing_2023}.

    \subsection{Fusion ordering}\label{sec:fusion-order}
    For a given fusion network $(H,F)$ we can also optimize the order in which we apply the fusions in $F$. Depending on the ordering, the mean first passage time until reaching the desired target graph state varies already for small instances (c.f. \cref{fig:fusion-order-independent}). Formally, the problem can be defined as follows: 
    \textit{Given a fusion network $(H,F)$ with target graph $G$, find an ordering in $F$, such that $M_{G\leftarrow H}$ is minimized.} 
    Finding the fusion order with minimal MFPT is challenging since the search space of possible orderings grows exponentially with the number of fusions and it is not immediately clear how the stochastic process evolves from the transition matrix. 
    
    Here we introduce a heuristic algorithm inspired from the \textit{min-weight-maximum-matching-first} algorithm introduced in~\cite{lee_graph-theoretical_2023}. We motivate this algorithm with the following observation: 
    Each intermediate state in the transition matrix has only two follow up states, namely, one on failure with probability $q$ and the other on success with probability $p$. The calculation of $M_{G\leftarrow H}$ can therefore be simplified to the following recursion, where $H$ is labeled as starting state $0$ and $G$ as target state $n$: \begin{align}
    	M_{n\leftarrow 0} &= 1 + qM_{n\leftarrow 0}+pM_{n\leftarrow 1}\\
    	M_{n\leftarrow i} &= 1 + qM_{n\leftarrow j}+pM_{n\leftarrow k}
    \end{align} For any intermediate state $i$, $j$ denotes the state being reached if the fusion at state $i$ fails and $k$ the state if the fusion succeeds. All $pM_{n\leftarrow k}$ terms are already fixed by fusion network and protocol, thus the only term we can optimize with different orderings is $qM_{n\leftarrow j}$. Assuming that $M_{n\leftarrow j} \geq M_{n\leftarrow i}$, i.e.: failed fusions always lead to states with higher mean first passage times than the current one, the best case for $qM_{n\leftarrow j}$ would be if $j=i$, that is, if on fusion failure we end up in the same state as before.

\begin{figure}
    	\begin{center}
        \resizebox{\linewidth}{!}{
    		\begin{tikzpicture}
	\begin{pgfonlayer}{nodelayer}
		\node [style=Cluster dot] (0) at (1.5, -0.25) {};
		\node [style=Cluster dot] (1) at (1.5, 0.75) {};
		\node [style=Cluster dot] (2) at (1.5, 1.75) {};
		\node [style=Cluster dot] (3) at (1.5, 2.75) {};
		\node [style=Cluster dot] (4) at (2.5, -0.25) {};
		\node [style=Cluster dot] (5) at (2.5, 2.75) {};
		\node [style=Cluster dot] (6) at (4.5, 2.75) {};
		\node [style=Cluster dot] (7) at (3.5, 2.75) {};
		\node [style=Cluster dot] (8) at (4.5, 0.75) {};
		\node [style=Cluster dot] (9) at (4.5, 1.75) {};
		\node [style=Cluster dot] (10) at (3.5, -0.25) {};
		\node [style=Cluster dot] (11) at (4.5, -0.25) {};
		\node [style=none] (12) at (5.9, 1.25) {$\rightarrow$};
		\node [style=Cluster dot] (13) at (7.15, 0.5) {};
		\node [style=Cluster dot] (14) at (7.15, 2) {};
		\node [style=Cluster dot] (15) at (9.65, 2) {};
		\node [style=Cluster dot] (16) at (9.65, 0.5) {};
		\node [style=Cluster dot] (17) at (10.9, 1.25) {};
		\node [style=none] (18) at (1, 1.25) {$a$};
		\node [style=none] (19) at (2, 3.5) {$b$};
		\node [style=none] (20) at (2, -1) {$c$};
		\node [style=none] (21) at (3, 3.4) {$e$};
		\node [style=none] (22) at (3, 1.25) {$d$};
		\node [style=none] (23) at (3, -1) {$f$};
		\node [style=none] (24) at (5, 2.25) {$g$};
		\node [style=none] (26) at (11.8, 1.25) {$\rightarrow$};
		\node [style=Cluster dot] (27) at (2.5, 0.75) {};
		\node [style=Cluster dot] (28) at (2.5, 1.75) {};
		\node [style=none] (29) at (5, 0.25) {$h$};
		\node [style=Cluster dot] (30) at (8.4, 0.5) {};
		\node [style=Cluster dot] (31) at (8.4, 2) {};
		\node [style=Cluster dot] (32) at (15.75, 1.25) {};
		\node [style=Cluster dot] (34) at (14.25, 0.5) {};
		\node [style=Cluster dot] (35) at (14.25, 2) {};
		\node [style=Cluster dot] (36) at (12.75, 1.25) {};
		\node [style=none] (47) at (2.75, -3.5) {$\rightarrow$};
		\node [style=Cluster dot] (48) at (4, -3.5) {};
		\node [style=none] (49) at (1.5, -3.5) {$\ldots$};
		\node [style=cluster] (50) at (6, -4.5) {$a$};
		\node [style=cluster] (51) at (7.5, -4.5) {$e$};
		\node [style=cluster] (53) at (6.75, -3.75) {$b$};
		\node [style=cluster] (54) at (9, -4.5) {$h$};
		\node [style=cluster] (55) at (8.25, -3.75) {$f$};
		\node [style=cluster] (56) at (7.5, -3) {$c$};
		\node [style=cluster] (57) at (8.25, -2.25) {$d$};
		\node [style=cluster] (58) at (9, -1.5) {$g$};
		\node [style=none] (59) at (10, -3.5) {$\rightarrow$};
		\node [style=none] (60) at (13.5, -3.5) {$[a,e,b,h,f,c,d,g]$};
		\node [style=none] (61) at (10, -2.75) {\small dfs};
		\node [style=none] (62) at (6.65, 1.25) {$a$};
		\node [style=none] (63) at (8.9, 2.5) {$e$};
		\node [style=none] (64) at (10.65, 0.5) {$h$};
		\node [style=none] (65) at (13.25, 2) {$b$};
		\node [style=none] (66) at (15.25, 0.5) {$f$};
	\end{pgfonlayer}
	\begin{pgfonlayer}{edgelayer}
		\draw (0) to (1);
		\draw (2) to (3);
		\draw (6) to (7);
		\draw (8) to (9);
		\draw (10) to (11);
		\draw [style=box edge] (1) to (2);
		\draw [style=box edge] (3) to (5);
		\draw [style=box edge] (4) to (0);
		\draw [style=box edge] (5) to (7);
		\draw [style=box edge] (6) to (9);
		\draw [style=box edge] (8) to (11);
		\draw [style=box edge] (10) to (4);
		\draw [style=hadamard edge] (13) to (14);
		\draw [style=box edge] (15) to (17);
		\draw [style=hadamard edge] (16) to (17);
		\draw [style=box edge] (27) to (28);
		\draw (5) to (28);
		\draw (27) to (4);
		\draw [style=box edge] (30) to (31);
		\draw [style=box edge] (13) to (30);
		\draw [style=box edge] (30) to (16);
		\draw [style=box edge] (31) to (14);
		\draw [style=hadamard edge] (31) to (15);
		\draw [style=box edge] (34) to (35);
		\draw [style=hadamard edge] (36) to (35);
		\draw [style=box edge] (34) to (36);
		\draw [style=box edge] (32) to (35);
		\draw [style=hadamard edge] (32) to (34);
		\draw (53) to (50);
		\draw (53) to (51);
		\draw (55) to (56);
		\draw (55) to (54);
		\draw (56) to (53);
		\draw (56) to (57);
		\draw (57) to (58);
	\end{pgfonlayer}
\end{tikzpicture}
            }
    	\end{center}
    	\caption{Example of finding a fusion order with the algorithm described in~\cref{sec:fusion-order}. The initial fusion network consists of eight fusions labeled from $a-h$. We repeatedly contract the network and find the maximal set of disjoint edges highlighted as dotted green edges until the network is contracted into a single node. At each iteration we add the disjoint fusions as vertices to a tree whose inverted dfs order determines the final fusion order.}
    	\label{fig:better-fusion-order-example}
    \end{figure}
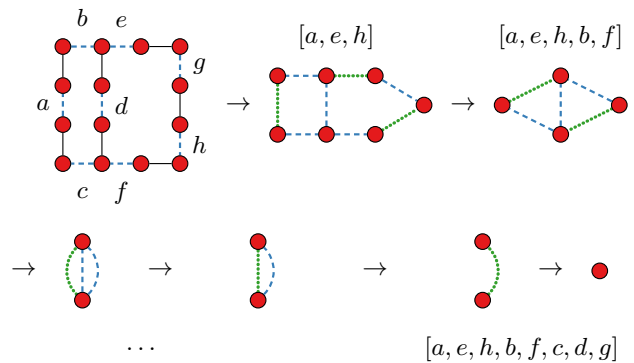

    This is the case if we apply fusions on disjoint resource states, because a failed fusion does not undo previous fusions and only destroys the involved resource states instead of qubits in a larger graph state. Since we assume that new resource states can be injected at any time into the creation process, we end up in the same state as before the fusion. The core idea of our algorithm thus is to choose an ordering where we apply fusions on disjoint connected sets of the graph as long as possible. For this we proceed as follows:
    \begin{figure}
    	\begin{center}
        \resizebox{\linewidth}{!}{
    		\begin{tikzpicture}
	\begin{pgfonlayer}{nodelayer}
		\node [style=Cluster dot] (0) at (-14, 3.5) {};
		\node [style=Cluster dot] (1) at (-13, 3.5) {};
		\node [style=Cluster dot] (2) at (-11.75, 3.5) {};
		\node [style=Cluster dot] (3) at (-10.75, 3.5) {};
		\node [style=Cluster dot] (4) at (-9.5, 3.5) {};
		\node [style=Cluster dot] (5) at (-8.5, 3.5) {};
		\node [style=Cluster dot] (6) at (-7.25, 3.5) {};
		\node [style=Cluster dot] (7) at (-6.25, 3.5) {};
		\node [style=none] (8) at (-12.35, 4) {$0$};
		\node [style=none] (9) at (-7.85, 4) {$1$};
		\node [style=none] (10) at (-10.1, 4) {$2$};
		\node [style=Cluster dot] (11) at (-13.25, 1) {};
		\node [style=Cluster dot] (12) at (-12.25, 1) {};
		\node [style=Cluster dot] (14) at (-11.25, 1) {};
		\node [style=Cluster dot] (15) at (-10, 1) {};
		\node [style=Cluster dot] (16) at (-9, 1) {};
		\node [style=Cluster dot] (17) at (-7.75, 1) {};
		\node [style=Cluster dot] (18) at (-6.75, 1) {};
		\node [style=Cluster dot] (35) at (-12.25, -1.75) {};
		\node [style=Cluster dot] (36) at (-11.25, -1.75) {};
		\node [style=Cluster dot] (37) at (-10.25, -1.75) {};
		\node [style=Cluster dot] (38) at (-9, -1.75) {};
		\node [style=Cluster dot] (40) at (-8, -1.75) {};
		\node [style=Cluster dot] (41) at (-7, -1.75) {};
		\node [style=Cluster dot] (42) at (-12, -4.5) {};
		\node [style=Cluster dot] (43) at (-11, -4.5) {};
		\node [style=Cluster dot] (44) at (-10, -4.5) {};
		\node [style=Cluster dot] (46) at (-9, -4.5) {};
		\node [style=Cluster dot] (47) at (-8, -4.5) {};
		\node [style=none] (48) at (-10.5, 2.75) {};
		\node [style=none] (49) at (-10.5, 1.5) {};
		\node [style=none] (53) at (-11.15, 2.1) {$p$};
		\node [style=none] (54) at (-10, 0.25) {};
		\node [style=none] (55) at (-10, -1) {};
		\node [style=none] (56) at (-6.5, 2.75) {};
		\node [style=none] (57) at (-7, -1) {};
		\node [style=none] (58) at (-5.6, 0.85) {$q$};
		\node [style=none] (59) at (-10.65, -0.4) {$p$};
		\node [style=none] (60) at (-10, -2.5) {};
		\node [style=none] (61) at (-10, -3.75) {};
		\node [style=none] (65) at (-10.65, -3.15) {$p$};
		\node [style=Cluster dot] (66) at (-4.225, 3.5) {};
		\node [style=Cluster dot] (67) at (-3.225, 3.5) {};
		\node [style=Cluster dot] (68) at (-1.975, 3.5) {};
		\node [style=Cluster dot] (69) at (-0.975, 3.5) {};
		\node [style=Cluster dot] (70) at (0.275, 3.5) {};
		\node [style=Cluster dot] (71) at (1.275, 3.5) {};
		\node [style=Cluster dot] (72) at (2.525, 3.5) {};
		\node [style=Cluster dot] (73) at (3.525, 3.5) {};
		\node [style=none] (74) at (-2.475, 4) {$0$};
		\node [style=none] (75) at (1.925, 4) {$2$};
		\node [style=none] (76) at (-0.325, 4) {$1$};
		\node [style=Cluster dot] (77) at (-3.475, 1) {};
		\node [style=Cluster dot] (78) at (-2.475, 1) {};
		\node [style=Cluster dot] (79) at (-1.475, 1) {};
		\node [style=Cluster dot] (80) at (-0.475, 1) {};
		\node [style=Cluster dot] (81) at (0.775, 1) {};
		\node [style=Cluster dot] (82) at (1.775, 1) {};
		\node [style=Cluster dot] (83) at (2.775, 1) {};
		\node [style=Cluster dot] (84) at (-2.725, -1.75) {};
		\node [style=Cluster dot] (85) at (-1.725, -1.75) {};
		\node [style=Cluster dot] (86) at (-0.725, -1.75) {};
		\node [style=Cluster dot] (87) at (0.275, -1.75) {};
		\node [style=Cluster dot] (88) at (1.275, -1.75) {};
		\node [style=Cluster dot] (89) at (2.275, -1.75) {};
		\node [style=Cluster dot] (90) at (-2.225, -4.5) {};
		\node [style=Cluster dot] (91) at (-1.225, -4.5) {};
		\node [style=Cluster dot] (92) at (-0.225, -4.5) {};
		\node [style=Cluster dot] (93) at (0.775, -4.5) {};
		\node [style=Cluster dot] (94) at (1.775, -4.5) {};
		\node [style=none] (95) at (-0.575, 2.75) {};
		\node [style=none] (96) at (-0.575, 1.5) {};
		\node [style=none] (97) at (0.175, 2.75) {};
		\node [style=none] (98) at (0.175, 1.5) {};
		\node [style=none] (99) at (0.825, 2.1) {$q$};
		\node [style=none] (100) at (-1.225, 2.1) {$p$};
		\node [style=none] (107) at (-0.725, -2.5) {};
		\node [style=none] (108) at (-0.725, -3.75) {};
		\node [style=none] (112) at (-1.375, -3.15) {$p$};
		\node [style=none] (113) at (-10.45, 4.5) {};
		\node [style=none] (114) at (-8.95, 4.5) {};
		\node [style=none] (115) at (-9.8, 5.85) {$q$};
		\node [style=none] (116) at (-0.975, 4.25) {};
		\node [style=none] (117) at (0.525, 4.25) {};
		\node [style=none] (118) at (-0.325, 5.6) {$q$};
		\node [style=none] (119) at (-9.85, 1.5) {};
		\node [style=none] (120) at (-8.35, 1.5) {};
		\node [style=none] (121) at (-9.2, 2.85) {$q$};
		\node [style=none] (125) at (-10, -5.5) {$M_{3\leftarrow 0}\approx10$};
		\node [style=none] (126) at (-0.225, -5.5) {$M_{3\leftarrow 0}\approx12$};
		\node [style=scalar] (127) at (-15.1, 3.5) {\textbf{0}};
		\node [style=scalar] (128) at (-15.1, 1) {\textbf{1}};
		\node [style=scalar] (129) at (-15.1, -1.75) {\textbf{2}};
		\node [style=scalar] (130) at (-15.1, -4.5) {\textbf{3}};
		\node [style=none] (131) at (-4.975, 6) {};
		\node [style=none] (132) at (-4.975, -6) {};
		\node [style=none] (133) at (-0.575, 0.25) {};
		\node [style=none] (134) at (-0.575, -1) {};
		\node [style=none] (135) at (0.175, 0.25) {};
		\node [style=none] (136) at (0.175, -1) {};
		\node [style=none] (137) at (0.825, -0.4) {$q$};
		\node [style=none] (138) at (-1.225, -0.4) {$p$};
	\end{pgfonlayer}
	\begin{pgfonlayer}{edgelayer}
		\draw (0) to (1);
		\draw (2) to (3);
		\draw (4) to (5);
		\draw (6) to (7);
		\draw [style=box edge] (5) to (6);
		\draw [style=box edge] (4) to (3);
		\draw [style=box edge] (2) to (1);
		\draw (11) to (12);
		\draw (15) to (16);
		\draw (17) to (18);
		\draw [style=box edge] (16) to (17);
		\draw [style=box edge] (15) to (14);
		\draw (14) to (12);
		\draw (35) to (36);
		\draw (40) to (41);
		\draw [style=box edge] (38) to (37);
		\draw (37) to (36);
		\draw (40) to (38);
		\draw (42) to (43);
		\draw (46) to (47);
		\draw (44) to (43);
		\draw (44) to (46);
		\draw [style=diredge, bend right] (48.center) to (49.center);
		\draw [style=diredge, bend right] (54.center) to (55.center);
		\draw [style=diredge, bend right] (57.center) to (56.center);
		\draw [style=diredge, bend right] (60.center) to (61.center);
		\draw (66) to (67);
		\draw (68) to (69);
		\draw (70) to (71);
		\draw (72) to (73);
		\draw [style=box edge] (71) to (72);
		\draw [style=box edge] (70) to (69);
		\draw [style=box edge] (68) to (67);
		\draw (77) to (78);
		\draw (80) to (81);
		\draw (82) to (83);
		\draw [style=box edge] (81) to (82);
		\draw [style=box edge] (80) to (79);
		\draw (79) to (78);
		\draw (84) to (85);
		\draw (88) to (89);
		\draw (87) to (86);
		\draw (86) to (85);
		\draw [style=box edge] (88) to (87);
		\draw (90) to (91);
		\draw (93) to (94);
		\draw (92) to (91);
		\draw (92) to (93);
		\draw [style=diredge, bend right] (95.center) to (96.center);
		\draw [style=diredge, bend right] (98.center) to (97.center);
		\draw [style=diredge, bend right] (107.center) to (108.center);
		\draw [style=diredge, bend left=255, looseness=2.25] (114.center) to (113.center);
		\draw [style=diredge, bend left=255, looseness=2.25] (117.center) to (116.center);
		\draw [style=diredge, bend left=255, looseness=2.25] (120.center) to (119.center);
		\draw [style=double edge] (131.center) to (132.center);
		\draw [style=diredge, bend right] (133.center) to (134.center);
		\draw [style=diredge, bend right] (136.center) to (135.center);
	\end{pgfonlayer}
\end{tikzpicture}
            }
    	\end{center}
        \caption{Transition graphs for two different fusion orders generating a five qubit linear graph and resp. mean first passage time for fusion success probability $p=0.5$. In the left example, the first and the second fusion are independent of each other, that is, they act on separate parts of the graph. Thus, the second fusion does not undo the first fusion which yields a better mean first passage time than for the right example.}
        \label{fig:fusion-order-independent}
    \end{figure}
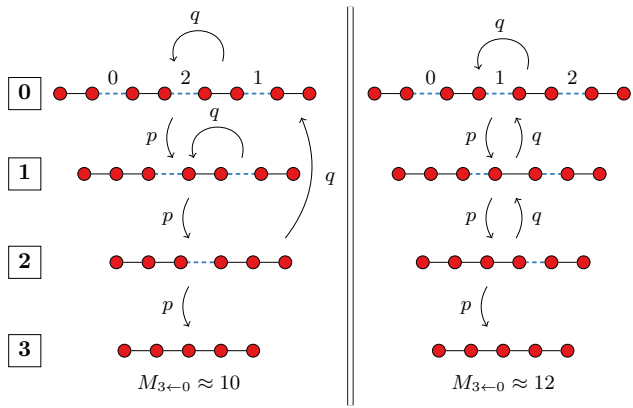
    \begin{enumerate}
    	\item We start with the initial fusion network and an empty tree $T$.
    	\item We contract the fusion network $(H,F)$ on all non fusion edges, that is all edges $E'\in E\setminus F$. This yields a graph $H'=(V,F)$ where each edge represents a fusion and each node a connected set of entangled qubits.
    	\item If $H'$ consists of only a single node we go to step 4, else we apply the Blossom algorithm~\cite{edmonds1965paths} on $H'$ to find a maximal set of disjoint edges $F'$. In our context we get a maximal set of fusions acting on disjoint connected sets of the graph. For each fusion we add a vertex to $T$: In the first iteration as leaves, in the subsequent iterations as internal vertices adjacent to all leaves of the corresponding connected sets. We then go back to step 2 with fusion network $(H',F\setminus F')$.
    	\item We return the fusion order as inverse dfs order on $T$ starting from its root vertex.
    \end{enumerate} 
    An illustrative example of this algorithm is shown in \cref{fig:better-fusion-order-example}.
    The standard implementation of the blossom algorithm runs in time $O(|E||V|^2)$, thus the entire fusion-order finding algorithm is upper bounded by $O(|F|^2|V|^2)$ for the contracted fusion network $H'=(V,F)$. 
    
    \section{Evaluation}\label{sec:evaluation}
    To analyze the improvement on mean first passage times with the developed algorithms we implemented a pipeline with the following steps: \begin{enumerate}
    	\item We optionally optimize the number of edges in the graph using either the tree decomposition method from Bouchet~\cite{bouchetTrees}, or the greedy heuristic in~\cite{staudacher_reducing_2023}.
    	\item We build a fusion network for the graph using the algorithms in~\cref{sec:generating-fusion-networks} and optionally optimize the fusion ordering.
    	\item We generate all possible transitions by running the corresponding protocol on the initial fusion network and iterate through all succeeding states in a depth-first way. Optionally after each fusion failure we again optimize the fusion order.
    	\item We build the transition matrix for a fixed fusion success probability and calculate the MFPT.
    \end{enumerate}
    We run this pipeline on both protocols for Type-I fusion with 2-qubit linear clusters and Type-II fusion with 3-qubit linear clusters as resource states. For each configuration, we average over ten graphs $G(m,n)$ that are randomly generated connected graphs with a fixed number of $m$ vertices and $n$ edges. The code used for the evaluations is available at Zenodo~\cite{staudacher_2026_18316338} allowing reproducibility and possible integration into other research projects.

    \subsection{Comparison against repeat until success scheme}
\begin{figure}
    	\resizebox{\linewidth}{!}{
    		\includegraphics{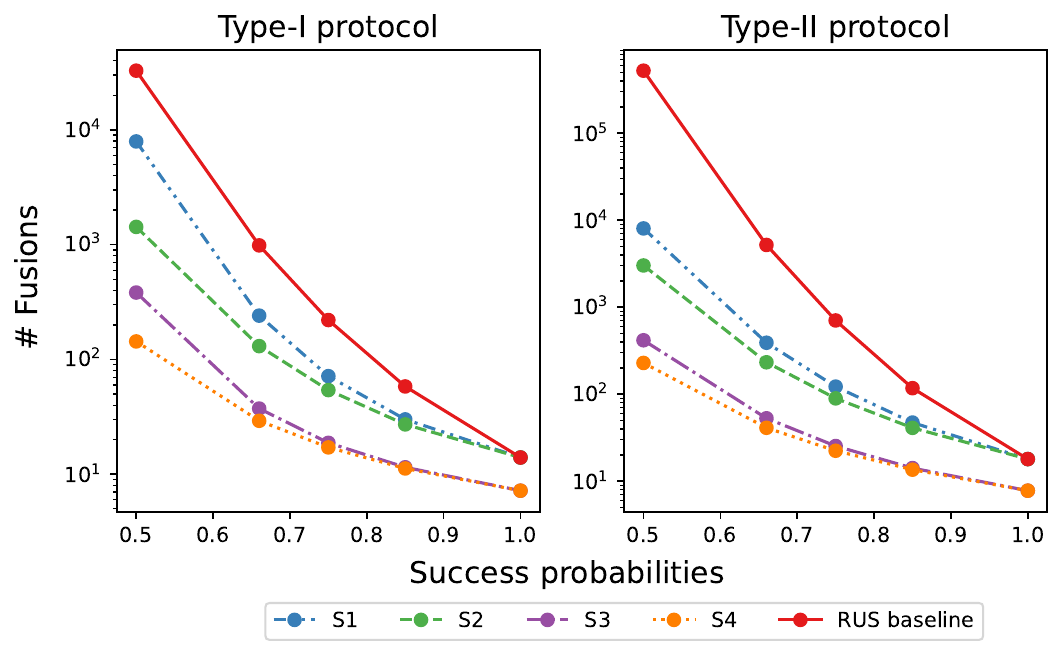}
    	}
    	\caption{Averaged mean first passage time for creating random graph states of 6 vertices and 10 edges with the strategies presented in~\cref{item:evalstrategies}. Additionally, we compare against a repeat-until-success baseline, that is, the mean first passage time, if we would restart the entire graph state building procedure anew after any fusion failure.}
    	\label{fig:6-10-mfpt-avg}
    \end{figure}

    \cref{fig:6-10-mfpt-avg} shows the averaged mean first passage times for random $G(6,10)$ graphs for varying fusion success probabilities 50, 66, 75 and 85\%. 
     We compare four different optimization strategies: \begin{itemize}\label{item:evalstrategies}
        \item[S1] The standard Type-I or Type-II fusion protocol without prior edge minimization via local complementation and no optimization of the fusion order.
        \item[S2] No edge minimization but with optimization of the fusion order.
        \item[S3] No optimization of the fusion order but with edge minimization.
        \item[S4] Both edge minimization and optimization of the fusion order.
    \end{itemize}
    We further compare against a repeat until success baseline where we restart the entire graph state buildup if any fusion fails.
    
    Our strategies clearly outperform the baseline for all fusion success probabilities which indicates that just using a protocol to recycle graph states on fusion failure improves the number of required fusions significantly. Both optimization of fusion orders and edge minimization via local complementation are then able to reduce the amount of required fusions by several orders of magnitude, especially for lower success probabilities. 
    In general, mean first passage times for the Type-II protocol are higher when compared to Type-I. One possible reason could be that in the Type-II protocol we discard more vertices from the graph state in the failure case. 
    
    \subsection{Comparison to Lee \& Jeong approach}
    In~\cite{lee_graph-theoretical_2023} authors propose a similar framework to ours for graph state generation from Type-II fusion measurements and 3-qubit linear clusters where they also first optimize edges in a graph via graph-theoretic rewrites, then extract fusion networks and last optimize the fusion order with the goal of minimizing the average number of fusions. Their protocol can be considered as an improved repeat-until-success scheme where fusions are arranged in a tree structure and the graph state is built in a bottom-up manner similar to Fig.~\ref{fig:better-fusion-order-example}. Each time a fusion fails, the graph state corresponding to the sub-tree below the failed tree node needs to be rebuilt entirely. In particular if the very last fusion fails, the buildup is reset to the beginning. With failed fusions on disjoint sub-trees not affecting each other their protocol poses a clear improvement over the naive repeat until success protocol from the previous section, yet it is still not adaptive in the sense of recycling leftover graph states. 
    \begin{figure}
    	\resizebox{.85\linewidth}{!}{
    		\includegraphics{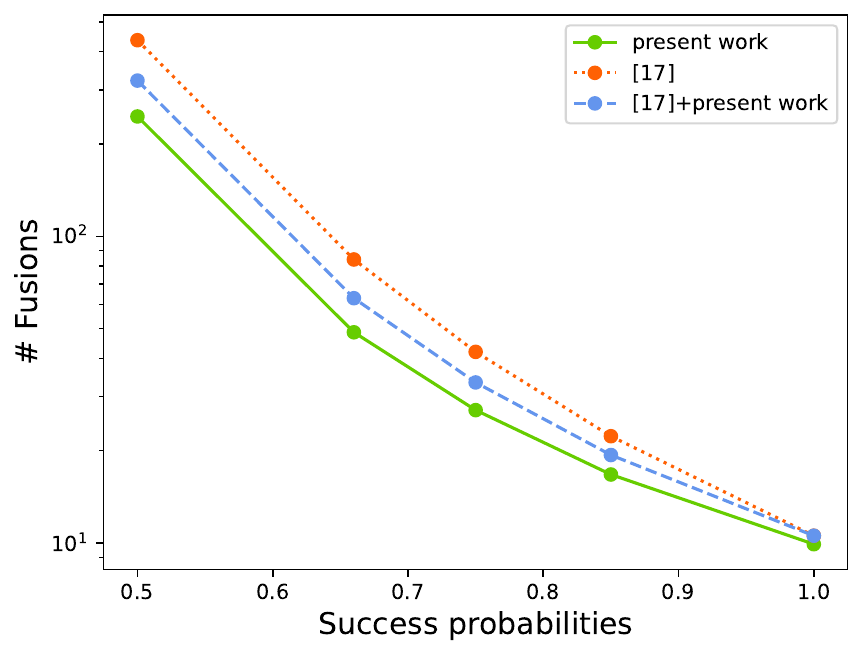}
    	}
    	\caption{Average expected number of fusions for $100$ connected random graphs with $V=|7|$ and $E=|12|$. \textit{Present work} corresponds to strategy S4, i.e. our compilation pipeline including both edge minimization and fusion order optimization, \textit{[17]} denotes the approach in~\cite{lee_graph-theoretical_2023}, and in \textit{[17]+present work} edge minimization and fusion network extraction are done with the approach in~\cite{lee_graph-theoretical_2023} and only the graph state buildup process is replaced by our adaptive protocol.}
    	\label{fig:7-12-leejeong}
    \end{figure}

    In Figure~\ref{fig:7-12-leejeong}, we compare strategy S4, i.e. our protocol with optimization of fusion order and prior edge minimization against the approach from~\cite{lee_graph-theoretical_2023} using the corresponding software package opt-graph-state~\footnote{The corresponding software package is available at\url{https://github.com/seokhyung-lee/OptGraphState}}. To better highlight the improvements achieved through adaptivity, we also include a strategy where edge minimization and fusion network extraction are done with the opt-graph-state package instead of our framework. 
    
    Results show that both strategies outperform the approach from~\cite{lee_graph-theoretical_2023}: 
    The combined strategy reduces the average number of fusions by up to 26\% when compared to~\cite{lee_graph-theoretical_2023} which clearly shows that adaptive protocols can reduce the expected number of fusions significantly. 
    With strategy S4 we obtain even higher reductions by up to 43\% indicating that edge minimization based on finding local equivalent graphs can reduce more edges than the edge minimization scheme in~\cite{lee_graph-theoretical_2023}.
    
    \subsection{Effect of fusion network optimizations}
\begin{figure*}
    \resizebox{\linewidth}{!}{
    	\includegraphics{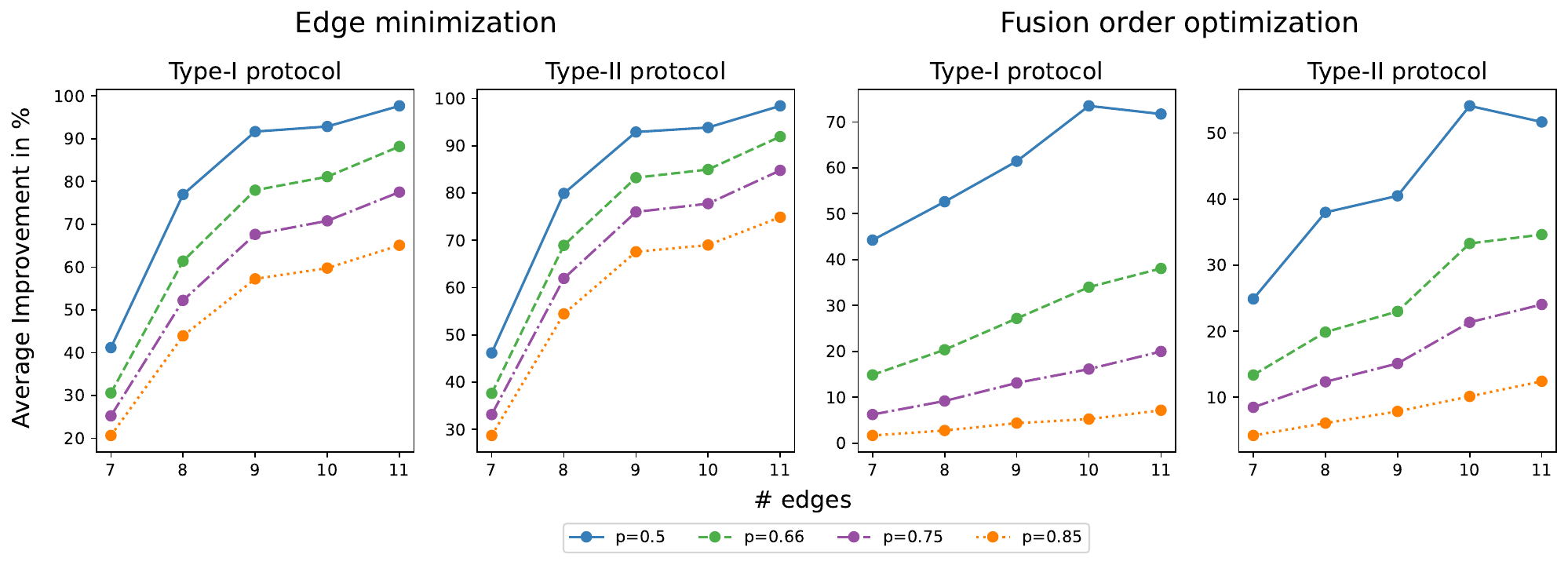}
        }
    	\caption{Averaged relative improvement of mean first passage times using the strategies described in \cref{sec:optimize-fusion-networks} for graphs of 6 nodes and varying number of edges. The two plots on the left show the improvement when minimizing edges in advance and the two plots on the right the improvement when optimizing the fusion order.}
    	\label{fig:6-x-lcomp-improvement}
    \end{figure*}

    We now analyze more in depth how different fusion network optimizations improve the mean first passage time.
    \cref{fig:6-x-lcomp-improvement} shows the relative improvement on the number of required fusions for graphs with and without prior edge minimization using local complementation on random graphs with $6$ nodes and increasing number of edges.
    While the average improvement is comparably small between $20-40$\% for graphs with $7$ edges, it increases continuously towards $70-100$\% in the case of $11$ edges. The smaller improvement for graphs with $7-8$ edges can be explained with the fact that in a connected graph with $m=6$ vertices there are not many edges to optimize since the number of edges is lower bounded by $m-1$, i.e.: the graph is a tree. In accordance to~\cref{fig:6-10-mfpt-avg} the highest optimization potential shows for protocols with lower success probabilities.
    
    In a similar way, \cref{fig:6-x-lcomp-improvement} shows the relative improvement on the number of required fusions when we optimize the fusion order during the protocol.
    Here the improvement is a bit more modest with an average improvement up to $20$\% for higher success probabilities and up to $70$\% for lower success probabilities. Still, results indicate that the more edges a graph has, the more improvement on mean first passage times is possible using the fusion order optimization.

\section{Conclusion and Outlook}
In this work, we proposed a framework to generate graph states with adaptive fusion protocols for both Type-I and Type-II fusion schemes on linear cluster resource states. In contrast to other work, we make use of adaptive fusion protocols allowing us to reprocess parts of leftover graph states after fusion failure has occurred. We utilized mean first passage times as a tool to estimate the number of required fusions in a sequential buildup process and investigated several algorithms to optimize fusion networks towards fewer fusion steps. Our results show substantial improvements in reducing the number of required fusions which can be up to several orders of magnitude when compared to simple repeat-until-success schemes. 

Our analysis can be extended in the direction of finding more efficient Type-II fusion networks by exploiting the pivot behavior of the $\{XZ,ZX\}$ basis measurement on graph states. Another line of exploration is to interleave different fusion types in the generation process.

Regarding speedups of the graph-state building process, we point out that an extension of our framework to incorporate parallel fusions is straightforward. It can be taken into account by redefining the states such that they correspond to concurrent manipulations of different parts of the graph. 

While in this work we focused only on all-photonic schemes, the Markov process formalism that we developed may also prove useful when analyzing other non-deterministic graph state generation protocols. Potential playgrounds are non-photonic cluster-states as they have been created for error-correction with robustness against single qubit errors~\cite{yao2012experimental}, graph states produced in trapped-ion systems~\cite{lanyon2013measurement} and superconducting platforms with 16 qubits~\cite{wang201816} and more recently 20 qubit systems on the IBM Q Poughkeepsie device \cite{mooney2019entanglement}. The past years have also seen a large number of developments regarding Rydberg atom systems, where graph states have also played a crucial role~\cite{bluvstein2022quantum}. All of these systems are currently subject to high levels of noise such that graph states have to be constructed in a stochastic fashion. Thus, formulating their construction as Markovian processes may turn out to be a fruitful perspective.

\section*{Acknowledgement}
The authors thank Francesco Debortoli for the early implementation of the translation to photonic circuits; Patrik Isene Sund for the valuable discussion on the design of optical fusions and quantum emitters as well as Francesco Giorgino and Lorenzo Carosini for discussion on experimental capabilities of such a scheme.
This research was funded in whole, or in part, from the Horizon Europe research and innovation programme under grant agreement No 101135288 (EPIQUE).

    \bibliographystyle{unsrt}
    \bibliography{sources}
     
    \newpage
    \appendix    
    
    \crefalias{section}{appendix}
    \crefalias{algorithm}{appalg}
    
    \renewcommand{\thealgorithm}{~\thesection.\arabic{algorithm}}
    \setcounter{algorithm}{0}
    
    \section{Derivation of the First Passage Time}
    \label{app:FirstPassage}
    In the main text we introduced the MFPT matrix $M_{ij}= M_{i\leftarrow j}$. This is the average number of steps to start
    at state $j$ and arrive at state $i$ for the first time. As we explained, the MFPT satisfies a simple recursion relation:
    \begin{equation}
    	M_{i\leftarrow j}=p_{i\leftarrow j}+\sum_{k\neq i}p_{k\leftarrow j}\left(M_{i\leftarrow k}+1\right).\label{eq:FirstPassage_eq}
    \end{equation}
    Thus, all we have to do in order to find our answer, is to solve this
    equation for $M$. We do this in several steps. First, we can simplify
    the \cref{eq:FirstPassage_eq}:
    \begin{align}
    	M_{i\leftarrow j} & =p_{i\leftarrow j}+\sum_{k\neq i}p_{k\leftarrow j}+\sum_{k\neq i}p_{k\leftarrow j}M_{i\leftarrow k}\nonumber
        \end{align}
    and thus
    \begin{align}
    	M_{i\leftarrow j} & =1+\sum_{k\neq i}p_{k\leftarrow j}M_{i\leftarrow k}.\label{eq:FPMain}
    \end{align}
    Secondly, we derive a useful fact. Multiply \cref{eq:FPMain}
    by the stationary vector component $\pi_{j}$ and sum over $j$:
    \[
    \sum_{j}M_{i\leftarrow j}\pi_{j}=\sum_{j}\pi_{j}+\sum_{j}\sum_{k\neq i}p_{k\leftarrow j}\pi_{j}M_{i\leftarrow k}.
    \]
    The definition of the stationary vector is that $\sum_{j}p_{k\leftarrow j}\pi_{j}=\pi_{k}$.
    Thus we have
    \[
    \sum_{j}M_{i\leftarrow j}\pi_{j}=1+\sum_{k\neq i}M_{i\leftarrow k}\pi_{k},
    \]
    where we also used $\sum_{j}\pi_{j}=1$, since $\pi$ is
    also a probability vector. Finally, moving the second term on the
    right to the left, we find that almost all terms cancel, except for
    one:
    \[
    M_{i\leftarrow i}\pi_{i}=1
    \]
    Thus we learn the diagonal entries of the MFPT matrix $M$:
    \[
    M_{i\leftarrow i}=\frac{1}{\pi_{i}}
    \]
    The next task is to identify the remaining entries of $M$. Starting
    from \cref{eq:FPMain}, we rewrite this as 
    \begin{align}
    	M_{ij} & =1+(M\cdot P)_{ij}-p_{i\leftarrow j}M_{i\leftarrow i}\\& =1+(M\cdot P)_{ij}-\frac{1}{\pi_{i}}p_{i\leftarrow j}\nonumber \\
    	\rightarrow M & =E+M\cdot P-D\cdot P\label{eq:FPMatForm}
    \end{align}
    where we defined the matrix with all entries equal to $1$:
    \[
    E_{ij}\equiv1
    \]
    And the diagonal matrix
    \[
    D_{ij}\equiv\frac{\delta_{ij}}{\pi_{i}}
    \]
    At this point, one may believe that the solution of \cref{eq:FPMatForm}
    is straightforward, since one just has to solve for $M$ by bringing the second term on the right to the left and multiply by $(1-P)^{-1}$. But this is not correct, since 1-P is singular and thus not invertible.
    It is singular because $P$ has an eigenvalue $1$. Thus $1-P$ has
    an eigenvalue $0$ and therefore $\det(1-P)=0$. However, it turns out that
    $E-D\cdot P$ is also singular. Nevertheless, it is possible to extract the matrix $M$. 
    
    \cref{eq:FPMatForm} can be solved differently.
    We begin by showing that the solution is unique. Assume \cref{eq:FPMatForm}.
    had another solution $M'$. Then 
    \begin{align*}
    	M & =E+M\cdot P-D\cdot P\\
    	M' & =E+M'\cdot P-D\cdot P
    \end{align*}
    Subtracting the equations, we obtain
    \[
    M-M'=(M-M')\cdot P.
    \]

    Interpreting this equation column by column, this equation states
    that each column of $M-M'$ is equal to the left eigenvector of $P$ corresponding
    to eigenvalue $1$. We already know that this eigenvector takes the form  $c\cdot(1,1,1,\dots1)$ for some real $c$.
    Moreover, we already proved that all $M$ matrices have diagonal elements
    $M_{ii}=1/\pi_{i}$. Thus $M-M'$ has zeros on the diagonal.
    But this implies $c=0$. Therefore
    \[
    M-M'=0,
    \]
    in other words the solution of \cref{eq:FPMatForm} is unique. 
    
    Then it remains to find this unique solution. We first state the solution and then prove that it solves \cref{eq:FPMatForm}. The solution is \textcompwordmark\begin{eqnarray}
    	\boxed{M=D(\bm{1}-Z+Z_{0}E) \label{eq:FSol}}.
    \end{eqnarray}Here, $Z$ is the fundamental matrix defined as
    \[
    Z=(\bm{1}-P+\Pi)^{-1},
    \] with $\Pi$ being the \textit{limiting matrix} of $P$. This is a matrix with the stationary $\pi$ vector repeated on $\dim{M}$ columns, i.e. $\Pi=(\pi,\pi,\dots,\pi)$.
    From this we can deduce the relation
    \begin{equation}
    	Z-Z\cdot P+\Pi=\bm{1},\label{eq:Zidentity}
    \end{equation}
    which will be needed below. The matrix $Z_{0}$ is the diagonal part
    of $Z$, i.e. $[Z_{0}]_{ii}=Z_{ii}$ and $[Z_{0}]_{ij}=0$ for all
    $i\neq j$. We can quickly checking that $M$ satisfies \cref{eq:FPMatForm}.
    First off, we can check that the diagonal parts of $M$ and $D$ match,
    as we found before:
    \[
    M_{ii}=\sum_{j}D_{ij}(\bm{1}-Z+Z_{0}E)_{ji}=D_{ii}(\bm{1}-Z+Z_{0}E)_{ii}
    \]
    and now notice that $-Z+Z_{0}E$ is zero on the diagonal. Thus the relation
    \[
    M_{ii}=D_{ii}
    \]
    holds. Next, we multiply $M$ in \cref{eq:FSol} from the right
    by $P$:
    \[
    M\cdot P=D(P-Z\cdot P+Z_{0}\cdot E)
    \]
    Here we used that $E\cdot P=E$ because the columns of $P$ sum to
    $1$. Now we use \cref{eq:Zidentity} and find
    \begin{align*}
    	M\cdot P & =D\cdot(P-\bm{1}-Z-\Pi+Z_{0}\cdot E)\\
    	&=M+D\cdot P - D\cdot \Pi \\
    	&=M+D\cdot P - E 
    \end{align*}
    where in the penultimate step we used \cref{eq:FSol} again. Thus we
    find that this $M$ satisfies \cref{eq:FPMatForm} and has diagonal
    entries $M_{ii}=1/\pi_{i}$ and we have therefore found
    the unique solution. 
    {\onecolumngrid
    	\section{Algorithms for fusion networks}
    	\label{app:Pseudocode} 
    	In this section we give pseudocode algorithms for graph state generation using Type-I fusion and two-qubit linear clusters following \cref{sec:graph-state-generation} and \cref{sec:generating-fusion-networks}. 
        \subsection{Graph State Generation Protocol}
    	\begin{algorithm}
    		\caption{Graph State Generation from Type-I fusion network}
    		\label{app:t1-graph-state-generation}
    		\begin{algorithmic}[1]
    			\Procedure{BuildGraphState}{$H$,$F$}
    			\While{$F\neq\emptyset$}
    			\State $v_1,v_2 \gets F$.pop()
    			\State success $\gets$ \Call{Fuse}{$v_1$,$v_2$}\Comment{Type-I fusion on photonic hardware}
    			\If{success} \Comment{Fusion success}
    			\ForAll{$n \in N(v_2)$}
    			\Call{AddEdge}{$H$,($v_1$,$n$)}
    			\EndFor
    			\State \Call{DeleteVertex}{$H$,$v_2$}
    			\Else \Comment{Fusion failure}
    			\State $H,F \gets$ \Call{UpdateFusionNetworkOnFailure}{$H$,$F$,$v_1$,$v_2$}
    			\EndIf
    			\EndWhile
    			\EndProcedure
    			\Procedure{UpdateNetworkOnFailure}{$H$,$F$,$v_1$,$v_2$}
    			\State $G \gets (\{v_1,v_2\},\emptyset)$ \Comment{Target graph for fusion sub-network}
    			\State $A \gets H.E\setminus F$ \Comment{Non-fusion edges}
    			\State $T \gets (H.V\setminus \{v_1,v_2\},\{(v,w)|(v,w)\in A \text{ and } (v,w \notin \{v_1,v_2\})\})$
    			\For{C $\in$ \Call{ConnectedComponents}{$T$}}
    			\State $ D \gets$ $\{(v,w)\in A | (v \in C) \text{ and } (w \in \{v_1,v_2\})\}$ \Comment{Collect destroyed edges}
    			\If{$|C| = 1$} \Comment{Replace component if it is too small}
    			\State $v,w \gets D$.pop()
    			\State $G\gets (G.V\cup \{v\},G.E\cup \{(v,w)\})$
    			\State \Call{DeleteVertex}{$T$,$v$}
    			\EndIf
    			\While{$D\neq\emptyset$} \Comment{Rebuild all destroyed edges}
    			\State $v,w \gets D$.pop()
    			\State $x\gets $\Call{AddNewVertex}{$G$}
    			\State $G\gets (G.V\cup \{x\},G.E\cup \{(w,x)\})$
    			\State $F = F\cup \{(v,x)\}$
    			\EndWhile
    			\EndFor
    			\State $S, F'\gets $ \Call{BuildFusionNetwork}{$G$} \Comment{Generate new fusion network}
    			\State $F \gets F \cup (v_1,v_2) \cup F'$ \Comment{Update fusions}
    			\State $H \gets (T.V\cup S.V,T.E\cup S.E\cup F)$ \Comment{Update fusion graph}
    			\State \Return $H,F$
    			\EndProcedure
    		\end{algorithmic}
    \end{algorithm}
    The \texttt{BuildGraphState} procedure sequentially iterates through all fusions of a given fusion network $(H,F)$ and at each step fuses the indicated qubits $v_1$ and $v_2$. Depending on the fusion outcome, the network is updated accordingly: In the success case, we process according to Definition~\ref{thm:t-1-fusion-network}, whereas in the failure case we call a separate procedure \texttt{UpdateNetworkOnFailure}. In here, we first create a graph $G$ containing $v_1$ and $v_2$ together with another graph $T$ representing the leftover graph state. We then reconstruct each edge between a vertex in $T$ and $v_1$ or $v_2$ since these edges got destroyed by fusion failure. To this end, we iterate through all the connected components $C$ in $T$ and collect, inside a new set $D$, all the destroyed edges that connect vertices in $C$ to  either $v_1$ or $v_2$. In the special case where $C$ consists only of a single vertex $v$, we remove it from the existing graph state $T$ and add the first edge in $D$ to $G$ where we label one endpoint with $v$. We then iterate over all remaining destroyed edges and add them to $G$ such that in the end $G$ contains all edges destroyed by fusion failure. Also during iteration we add fusions to connect the endpoints of the destroyed edges to the leftover graph state.
    We then build a fusion network for $G$ and merge it with the existing network where we additionally add the fusion $\{v_1,v_2\}$ which failed originally. 
    \subsection{Fusion Network Generation}
    \begin{algorithm}
    	\caption{Type-I random traversal fusion network generation}
    	\label{app:RandomNetwork}
    	\begin{algorithmic}[1]
    		\Procedure{BuildFusionNetwork}{$G$}
    		\State $H \gets (V=\emptyset,E=\emptyset)$ \Comment{Init fusion network}
    		\State $F \gets \emptyset$ \Comment{Additional fusions}
    		\State $P\gets \emptyset$ \Comment{Position map}
    		\State $E\gets$ randomShuffle$(G.E)$ \Comment{Traverse target graph in random order}
    		\While{$E\neq \emptyset$}
    		\State $v,w \gets E$.pop()
    		\State $x_1,x_2\gets$\Call{AddVertices}{$H$} \Comment{Add resource state to network}
    		\State $H.E\gets H.E\cup \{x_1,x_2\}$
    		\If{$v\in P\text{ and } w \in P$} \Comment{Fuse resource state with existing network}
    		\State $F\gets F\cup\{(x_1,P[v]),(x_2,P[w])\}$
    		\ElsIf{$v\in P\text{ and } w \notin P$}
    		\State $F\gets F\cup\{(x_1,P[v])\}$
    		\ElsIf{$v\notin P\text{ and } w \in P$}
    		\State $F\gets F\cup\{(x_2,P[w])\}$
    		\EndIf
    		\State $P[v] \gets x_1$ \Comment{Update position map}
    		\State $P[w] \gets x_2$
    		\EndWhile
    		\State $H.E\gets H.E\cup F$
    		\State \Return $H,F$
    		\EndProcedure
    	\end{algorithmic}
    \end{algorithm}
    The \texttt{BuildFusionNetwork} procedure first initializes an empty fusion network $(H,F)$ and a position map to store which vertex in the original graph $G$ corresponds to which vertex in the fusion network. For each edge $\{v,w\}$ in $G$ we add a new two-qubit linear cluster to $H$. Depending on the three cases outlined in \cref{fig:tree-vs-loop-edges} we add fusions to $F$ connecting the new cluster with the existing fusion network. Further, the position map is updated where the new cluster now represents the edge $\{v,w\}$. At the end we add the edges in $F$ to $H$ and return the fusion network. 
}
\end{document}